%% file: main.tex
\documentclass[letterpaper,journal]{IEEEtran}
\usepackage{amsmath,amsfonts}
\usepackage{algorithmic}
\usepackage{array}
\usepackage[caption=false,font=normalsize,labelfont=sf,textfont=sf]{subfig}
\usepackage{textcomp}
\usepackage{stfloats}
\usepackage{url}
\usepackage{verbatim}
\usepackage{graphicx}
\usepackage{cite}
\usepackage{amsmath, amssymb, amsfonts}

\usepackage{graphicx}
\usepackage{xcolor}

\usepackage{booktabs}
\usepackage{makecell}
\usepackage{multirow}
\usepackage{adjustbox}
\usepackage{threeparttable}
\usepackage{arydshln}

\usepackage{float}
\usepackage{subfig}
\usepackage{subcaption}

\usepackage{pifont}
\usepackage{soul}
\usepackage{ragged2e}

\usepackage{enumitem}

\usepackage{url}

\usepackage[ruled,vlined,linesnumbered]{algorithm2e}

\definecolor{MyBlueGreen}{rgb}{0.0, 0.87, 0.87}
\definecolor{MyRed}{rgb}{0.9, 0, 0}
\definecolor{MyGreen}{rgb}{0, 0.9, 0.45}
\definecolor{MyBlack}{rgb}{0, 0, 0}

\newcommand{\eg}{\textit{e.g.}}
\newcommand{\etal}{\textit{et al.}}
\newcommand{\etc}{\textit{etc}}

\newcommand\zcomment[1]{\textcolor{MyBlack}{#1}}
\newcommand\hcomment[1]{\textcolor{MyBlack}{#1}}
\newcommand\gcomment[1]{\textcolor{MyBlack}{#1}}

\begin{document}

\title{SyncGait: Robust Long-Distance Authentication for Drone Delivery via Implicit Gait Behaviors}

\author{
Zijian~Ling, 
Man~Zhou,~\IEEEmembership{Member, IEEE}, 
Hongda~Zhai, 
Yating~Huang,
Lingchen~Zhao, \\
Qi~Li,~\IEEEmembership{Senior Member, IEEE}, 
Chao~Shen,~\IEEEmembership{Senior Member, IEEE}, 
and~Qian~Wang,~\IEEEmembership{Fellow, IEEE}

\IEEEcompsocitemizethanks{\IEEEcompsocthanksitem Z. Ling, M. Zhou, H. Zhai, and Y. Huang are with Hubei Key Laboratory of Distributed System Security, Hubei Engineering Research Center on Big Data Security, School of Cyber Science and Engineering, Huazhong University of Science and Technology, Wuhan 430074, China (e-mail: \{zijianling, zhouman, zhd, huangyating\}@hust.edu.cn).
\IEEEcompsocthanksitem L. Zhao and Q. Wang are with the Key Laboratory of Aerospace Information Security and Trusted Computing, Ministry of Education, School of Cyber Science and Engineering, Wuhan University, Wuhan 430072, China (e-mail: \{lczhaocs, qianwang\}@whu.edu.cn).
\IEEEcompsocthanksitem Q. Li is with the Institute for Network Sciences and Cyberspace, Tsinghua University, Beijing 100084, China (e-mail: qli01@tsinghua.edu.cn).
\IEEEcompsocthanksitem C. Shen is with the MOE Key Laboratory for Intelligent Networks and Network Security and the School of Cyber Science and Engineering, Xi'an Jiaotong University, Xi'an 710049, China (e-mail: chaoshen@mail.xjtu.edu.cn).
}

}

\maketitle

\begin{abstract}
In recent years, drone delivery, which utilizes unmanned aerial vehicles (UAVs) for package delivery and pickup, has gradually emerged as a crucial method in logistics.
Since delivery drones are expensive and may carry valuable packages, they must maintain a safe distance from individuals until user-drone mutual authentication is confirmed. Despite numerous authentication schemes being developed, existing solutions are limited in authentication distance and lack resilience against sophisticated attacks.
To this end, we introduce SyncGait, an implicit gait-based mutual authentication system for drone delivery. SyncGait leverages the user's unique arm swing as he walks toward the drone to achieve mutual authentication without requiring additional hardware or specific authentication actions.
We conducted extensive experiments on 14 datasets collected from 31 subjects. The results demonstrate that SyncGait achieves an average accuracy of 99.84\% at a long distance ($>18m$) and exhibits strong resilience against various spoofing attacks, making it a robust, secure, and user-friendly solution in real-world scenarios.
\end{abstract}


\input{Section/1_Introduction}

\input{Section/2_Related_Work}

\input{Section/3_Threat_Model}

\input{Section/4_System_Design_revised}

\input{Section/6_Evaluation_revised}

\section{Conclusion}
In this paper, we propose SyncGait, a mutual user–drone authentication system based on implicit gait behaviors. It utilizes the user’s natural arm swing during approach to the drone, requiring no additional hardware or explicit actions. Extensive outdoor experiments across 14 datasets demonstrate an average accuracy of 99.84\%, confirming its robustness and practicality. SyncGait effectively defends against radio relay, device hijacking, and mimicry attacks, representing the first user–drone authentication system resilient to device hijacking and supporting long-distance ($>18$ m) authentication.


\bibliographystyle{IEEEtran}
\bibliography{Section/Sample}

\end{document}

%% file: Section/1_Introduction.tex
\section{Introduction}
With the rapid advancement of UAV technology, drone delivery is gradually becoming an important method in the logistics industry. Currently, many companies, \eg, Amazon Air~\cite{Amazon-Air}, UPS~\cite{UPS}, and DroneUp~\cite{DroneUp}, are promoting drone delivery. 
According to BCC Research, the global market for drone delivery has reached \$1.51 billion in 2024 and will maintain a compound annual growth rate (CAGR) of 31.3\% in the next five years~\cite{bcc2024}.

Although this disruptive service can bring great convenience, it also introduces new challenges for the security and privacy of package delivery and pickup. Without effective mechanisms to verify the identities of users and drones, attackers may impersonate recipients to intercept packages illegally in drone-based delivery services. Similarly, in drone-based pickup services, attackers could disguise their drones as legitimate ones to steal packages. 
Given that delivery drones are often expensive and may carry valuable packages, there is a significant risk that they could be maliciously shot down if they approach residents before confirming their identities~\cite{qz2017}. 
Therefore, developing a secure and reliable long-distance user-drone mutual authentication mechanism has become an urgent priority for advancing the drone logistics industry.

Early drone authentication schemes~\cite{shucker2016machine,ganesh2016methods,Singireddy2018,sounduav} primarily focused on user authentication, \eg, 
Google proposed a solution that verifies the identity by scanning a QR code displayed on the user's smartphone~\cite{shucker2016machine}. 
Amazon's Prime Air project equips drones with biometric scanners, including fingerprint, retina, and facial recognition, to verify the recipient’s identity~\cite{Singireddy2018}. 
These schemes only authenticate the user, neglecting the possibility of a malicious drone impersonating a legitimate delivery drone to steal packages. 
To avoid the security risk of unilateral authentication, some mutual authentication schemes~\cite{natarajan2020unmanned,smile2auth,g2auth,h2auth,zhang2025droneaudioid} for drone delivery have been proposed recently. 
Users and drones can authenticate each other through Bluetooth beacon signals~\cite{natarajan2020unmanned}, Distance bounding~\cite{avoine2018security}, specially designed phone waving~\cite{g2auth}, facial expressions~\cite{smile2auth}, unique drone noises~\cite{h2auth}, and the drone's acoustic fingerprint~\cite{zhang2025droneaudioid}.
These solutions have provided new strategies for enhancing the security of drone delivery. However, several issues remain to be resolved: 
1) \emph{Inability to support long-distance authentication.}
Many solutions rely on high-resolution detailed features. 
This reliance results in poor performance at long distances, with typical authentication distances of $3\sim5 m$, 
making drones vulnerable to being shot down. 
\gcomment{2) \emph{Weak resistance to sophisticated attacks.}
These schemes assume that the user's phone is bound to the user, making it possible for an attacker to bypass the authentication mechanism by physically or digitally hijacking the user's smartphone. 
}
\gcomment{3) \emph{High demands on user compliance.} Current methods usually require prolonged authentication durations or explicit authentication actions. }

\begin{table*}[t!]
\caption{Comparison of user-drone authentication schemes.}
\label{tab:auth_comparison}
\vspace{-5pt}
\renewcommand{\scriptsize}{\fontsize{7.8}{10}\selectfont}
\scriptsize
\setlength{\tabcolsep}{2.5pt}
\begin{minipage}{\textwidth}

\begin{center}
{
\begin{tabular}{l c| c c c c | c c c}
    \hline\rule{0pt}{12pt}
    Scheme
    & \makecell{Mutual \\ Authentication}
    & \makecell{No Extra \\Hardware}
    & \makecell{No explicit \\Auth Actions} 
    & \makecell{Ambient \\Robustness$^1$}
    & \makecell{Auth Horizontal \\Distance}
    & \makecell{Resists Radio \\Relay Attacks}
    & \makecell{Resists Mimicry \\Attacks}
    & \makecell{Resists Device \\Hijacking} 

    \\ 
    \hline\rule{0pt}{8pt}

    Face/fingerprint recognition~\cite{Singireddy2018}
    & \ding{55}
    & \ding{55}
    & \ding{55}
    & \ding{55}
    & $<1m$
    & \ding{55}
    & \ding{55}
    & \ding{55}
    \\

    Google (QR code)~\cite{shucker2016machine}
    & \ding{55}

    & \ding{55}
    & \ding{55}
    & \ding{55}
    & $<1m$
    & \ding{55}
    & \ding{55}
    & \ding{55}
    \\
    
    Walmart (Bluetooth beacon)~\cite{natarajan2020unmanned}
    & \ding{51}
    & \ding{55}
    & \ding{51} 
    & \ding{51} 
    & $\sim5m$
    & \ding{55}
    & \ding{55}
    & \ding{55}
    \\

    Distance bounding~\cite{avoine2018security}
    & \ding{51}
    & \ding{51} 
    & \ding{51} 
    & \ding{55} 
    & $\sim5m$
    & \ding{51}
    & \ding{55}
    & \ding{55}
    \\

    SoundUAV~\cite{sounduav}
    & \ding{55}

    & \ding{55}
    & \ding{51}
    & \ding{55}
    & $\sim5m$
    
    & \ding{55}
    & \ding{51}
    & \ding{55}
    \\
    
    G2Auth~\cite{g2auth} 
    & \ding{51}
    & \ding{51} 
    & \ding{55} 
    & \ding{51} 
    & $5m$ 
    & \ding{51}
    & \ding{51}
    & \ding{55}
     \\

    Smile2Auth~\cite{smile2auth} 
    & \ding{51}
    & \ding{51} 
    & \ding{55} 
    & \ding{55}
    & $\sim3m$
    & \ding{51}
    & \ding{51}
    & \ding{55}
     \\

    H2Auth~\cite{h2auth} 
    & \ding{51}
    & \ding{51} 
    & \ding{51} 
    & \ding{55} 
    & $5m$ 
    & \ding{51}
    & \ding{51}
    & \ding{55}
     \\

    DroneAudioID~\cite{zhang2025droneaudioid}
    & \ding{51}
    & \ding{51} 
    & \ding{51} 
    & \ding{51} 
    & $2.4m$ 
    & \ding{51}
    & \ding{51}
    & \ding{55}
    \\
    \hline

    \textbf{SyncGait} 
    & \ding{51}
    & \ding{51} 
    & \ding{51} 
    & \ding{51} 
    & $\mathbf{>18m}$
    & \ding{51}
    & \ding{51}
    & \ding{51}
     \\
    \hline
\end{tabular}
}
\end{center}
    \footnotesize
    $^1$ Maintain reliability under varying environmental conditions, \eg, lighting and noises. 
\vspace{-5pt}
\end{minipage}
\end{table*}

In this paper, we propose SyncGait, a novel user-drone mutual authentication system based on implicit gait behaviors. During the drone delivery process, SyncGait utilizes the unique arm swing associated with the gait features during the user walking toward the drone, enabling mutual authentication between the drone and the user. 
Human gait characteristics have been demonstrated to be unique and difficult to imitate
~\cite{eberz2018your}. 
SyncGait leverages the temporal consistency between users and delivery drones, as well as the unforgeable nature of arm swing, to achieve long-distance, secure, and user-friendly mutual authentication.
Table \ref{tab:auth_comparison} compares SyncGait with existing user-drone mutual authentication schemes. SyncGait has several distinct advantages: 
1) \emph{Supporting long-distance authentication.} 
\zcomment{SyncGait extracts hand movement features from the user’s hand contours rather than relying on fine-grained details,} 
which facilitates long-distance user authentication. 
2) \emph{No extra hardware required.} SyncGait leverages the inertial measurement unit (IMU) sensors of smartphones and drone cameras for hand motion data collection, incurring no extra hardware costs. 
3) \emph{Implicit authentication.} 
Authentication is completed seamlessly as users walk toward the drone with a smartphone in hand, requiring no explicit interaction and imposing no additional burden on the user.
4) \emph{Security and reliability.} Gait is an inherent physical characteristic influenced by muscles, bones, and behavioral habits, making it difficult to forge or imitate. The arm swing during the user's walking provides implicit gait information~\cite{yang2023irga}.

SyncGait presents significant potential for widespread application but still faces several technical challenges: 1) The hand movements captured by the phone's IMU reflect the hand's positional changes in three-dimensional space, while the corresponding video primarily captures these changes through depth information. Although recent studies~\cite{yuan2022neural,yang2024depth} have attempted to extract depth information using deep learning, these methods often suffer from low accuracy and extensive computing resources.
2) IMU data provides absolute spatial motion information, while video data reflects relative motion between various body parts. The discrepancy between these modalities, compounded by the varying speed of hand movement during walking, complicates the process of temporal-spatial consistency analysis. 
3) 
As the user walks towards the drone, the vertical angle continuously increases.
High vertical angles would result in occlusion or distortion of key body features. The significant variations in gait sequences captured from different vertical angles inevitably raise the complexity of recognition.

To address these challenges, SyncGait introduces several innovative techniques:  
To analyze the periodic changes in three-dimensional hand posture, SyncGait leverages keypoint correction on the body skeleton to account for changes in coordinate systems over time, along with normalization to ensure the consistency of hand motion cycle data. 
Additionally, 
through the proposed Gradient-Aided AHRS Algorithm, 
SyncGait transforms IMU data from the phone's spatial coordinates to the Earth's spatial coordinates. 
\zcomment{An adaptive Butterworth filter dynamically adjusts its cutoff frequencies based on spectral energy distribution analysis of gait-related keypoint trajectories, effectively minimizing errors induced by individual user-specific behaviors.}
Subsequently, we convert the velocity vector in the earth's spatial coordinate to the human body posture-relative coordinate consistent with the video data, achieving spatial consistency between the two data, and ensuring their temporal consistency through our proposed temporal transformation algorithm. 
Moreover, observing that the occlusion of the human body under high vertical angles is only short-lived, a posture calibration algorithm is proposed to adaptively filter out anomalies, effectively mitigating occlusion impacts across varying viewpoints.
Our contributions are as follows:
\begin{itemize}
\item We present SyncGait, the first system that enables secure and robust long-distance ($>18m$) mutual authentication during the user walking towards the drone. SyncGait requires no additional hardware or explicit user authentication actions, operating seamlessly by using the smartphone’s IMU and the drone’s onboard camera.

\item We propose a posture calibration algorithm to mitigate body distortion and deformation caused by vertical-angle occlusion and long-distance blurring. In addition, we propose a cross-modal spatiotemporal preprocessing framework that effectively resolves temporal and spatial inconsistencies between visual and inertial modalities.

\item We prototype SyncGait on various commercial devices, and conduct extensive experiments to evaluate its performance under different settings, including challenging conditions, \eg, low lighting, occlusion of body parts, complex environment, data packet loss, \etc. The results show that SyncGait achieves an average accuracy of 99.84\% at 0.09\% EER, demonstrating strong robustness and generalizability for practical deployment.

\item We evaluate the security of SyncGait in defeating various spoofing attacks, including radio relay attack, device hijacking attack, and mimicry attack.
Experiment results show that attack success rates are all below 1\%.
\end{itemize}

%% file: Section/2_Related_Work.tex
\section{Related Work}

\subsection{User-Unmanned Devices Authentication.} 
In existing studies on authentication between users and unmanned devices (\eg, drones and autonomous vehicles), most methods rely on comparing the correlation between data from the device and the user's end. This approach has been widely applied in~\cite{g2auth, smile2auth, w4s, sounduav, h2auth, wu2022use, delgado2024act2auth,zhang2025droneaudioid}.

Some studies focus on user-specific characteristics for authentication. 
For instance, G2Auth~\cite{g2auth} 
\hcomment{authenticates users by comparing the movement trajectory of the phone, calculated using data from phone IMU sensors and video captured by the drone. However, this process is easy to imitate. MA-trained mimicry attacks achieved a significantly higher attack success rate of up to 26\% against G2Auth.} 
Similarly, Smile2Auth~\cite{smile2auth} relies on comparing simultaneous smile videos captured by both the user and the drone for authentication. This method relies on high-definition facial images,  yet even with the drone recording in 2.7K resolution at a horizontal distance of 3.3m, the accuracy is only 86.5\%. 
All of these methods~\cite{g2auth, smile2auth} require users to perform extra actions (such as hand-waving or smiling) and fail to meet the demands of long-distance authentication.
Additionally, some methods leverage the unique characteristics of the drone itself. SoundUAV~\cite{sounduav} authenticates by recognizing the distinctive motor noises of each drone. 
H2Auth~\cite{h2auth} authenticates by comparing the drone noises recorded by both the drone and the smartphone. Nevertheless, this method struggles to resist high-energy attacks and outdoor wind noises, making it unsuitable for long-distance authentication. 
DroneAudioID~\cite{zhang2025droneaudioid} authenticates by using drones' unique acoustic fingerprints extracted from fundamental and harmonic frequency components, but it lacks user-side identity binding and shows limited robustness, with a maximum tested authentication distance of only 2.4 m and no defense against advanced physical-layer attacks.
Due to the limited authentication distance, these solutions leave drones highly susceptible to malicious shooting down.
In contrast, SyncGait eliminates the need for users to rely on additional hardware or perform extra actions. Users simply need to carry a smartphone and walk normally, enabling reliable mutual authentication across various environmental conditions. Notably,  SyncGait maintains exceptional robustness even in long-distance ($>$18m) authentication, making it more secure and practical for real-world applications.

\subsection{IMU-based Gait Authentication.} 
Traditional gait authentication approaches using inertial sensors have primarily relied on template matching techniques~\cite{ren2014user,muaaz2017smartphone,alobaidi2022real}. Recently, gait authentication methods leveraging deep learning have shown significantly superior performance~\cite{chao2021gaitset,9714177,filipi2022gait,zou2020deep,delgado2023exploring,sezavar2024dcapsnet,yang2023irga,tran2022security}.
Previous research has extensively studied the application of deep learning models as feature extractors to derive gait characteristics from raw data~\cite{tran2022security,sezavar2024dcapsnet,delgado2023exploring}. 
Tran~\etal~\cite{tran2022security} employed a convolutional neural network (CNN) to extract gait features, which were then combined with a support vector machine (SVM) for user authentication. 
Sezavar~\etal~\cite{sezavar2024dcapsnet} extracted local features through convolutional layers and utilized a deep capsule network to capture the spatial relationships between these features. 
In~\cite{delgado2023exploring}, the Transformer architecture was applied to IMU-based gait authentication, aiming to effectively capture global contextual features through the self-attention mechanism.
In recent years, hybrid models have gained considerable attention in IMU-based gait authentication~\cite{zou2020deep,yang2023irga}. 
Zou~\etal~\cite{zou2020deep} proposed a hybrid model combining long short-term memory (LSTM) networks with CNNs, employing both as feature extractors to process time-series data. Yang~\etal~\cite{yang2023irga} extracted spatial features from IMU data using convolutional layers, and then fed these into an LSTM layer to learn temporal features.
Different from the aforementioned works, our method incorporates magnetometer data for correcting rotational angles in deep learning-based characterizations of gait behaviors, utilizing a hybrid model based on ResNet and LSTM for feature extraction. 

%% file: Section/3_Threat_Model.tex
\section{Preliminary}

\begin{figure}[t]
  \centering
  \includegraphics[width=1.0\linewidth]{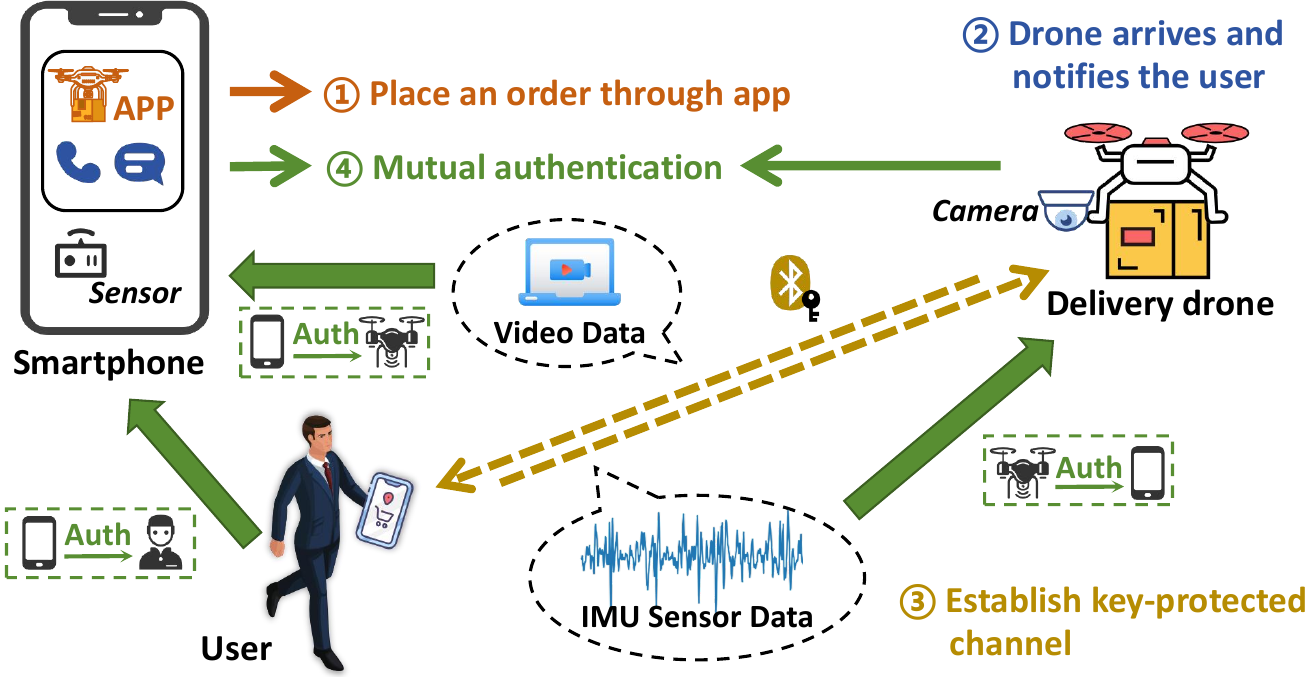} 
  \vspace{3pt}
  \caption{\zcomment{The authentication process of SyncGait.}}
  \label{fig:process} 
\end{figure}

\subsection{Authentication Process of SyncGait}
As illustrated in Figure~\ref{fig:process}, the authentication process of SyncGait is detailed as follows:
\begin{enumerate}[leftmargin=2em, topsep=0pt, partopsep=0pt, itemsep=0pt, parsep=0pt]
\renewcommand{\labelenumi}{(\theenumi)}
    \item The user places an order through a delivery app on his/her phone. After arriving at the designated location, the drone hovers in the air and notifies the user of its arrival through SMS, phone calls, \etc. Then, it establishes a key-protected \zcomment{Bluetooth} channel with the user's phone.
    
    
    \item The user holds the phone and walks toward the designated location. The drone confirms the user’s real-time position through shared GPS data and searches for the user using its camera. Once the user is detected, the drone initiates the mutual authentication process.

    \item During this process, the drone captures the user's video and processes it in real-time to extract time-series of the user’s body keypoints. Meanwhile, the user’s phone continuously collects time-series data from its IMU sensors. These two data are then exchanged between the drone and the user's phone.

    \item 
    \zcomment{During the \textit{Consistency Verification}, the drone authenticates the user by comparing the IMU data received from the user’s phone with its locally captured video data, while the user’s phone verifies the drone by comparing the exchanged video data with its local IMU data. Meanwhile, the phone conducts \textit{IMU-based Gait Authentication} based on the local IMU data.}
    If all determinations are passed, the mutual authentication succeeds. Otherwise, the drone determines whether it needs to adjust its position based on the distance from the user, and the system returns to Step 3, repeating the process until the maximum number of attempts is reached.
\end{enumerate}

\begin{figure}[t]
  \centering
  \includegraphics[width=1\linewidth]{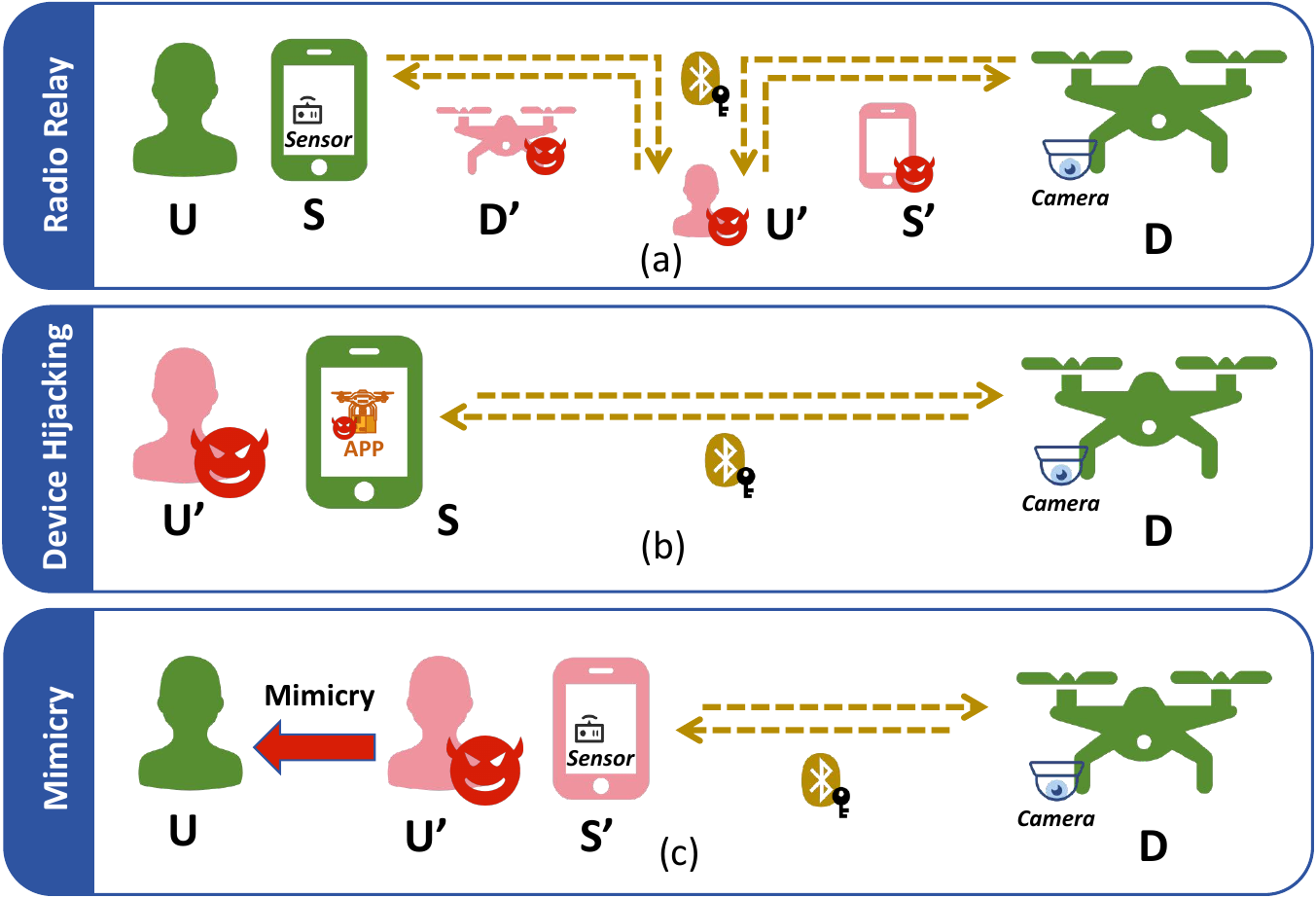} 
  \vspace{3pt}
  \caption{\zcomment{Different types of attacks.}}
  \label{fig:attacks} 
\end{figure}

\begin{figure*}[t]
  \centering
  \includegraphics[width=1\linewidth]{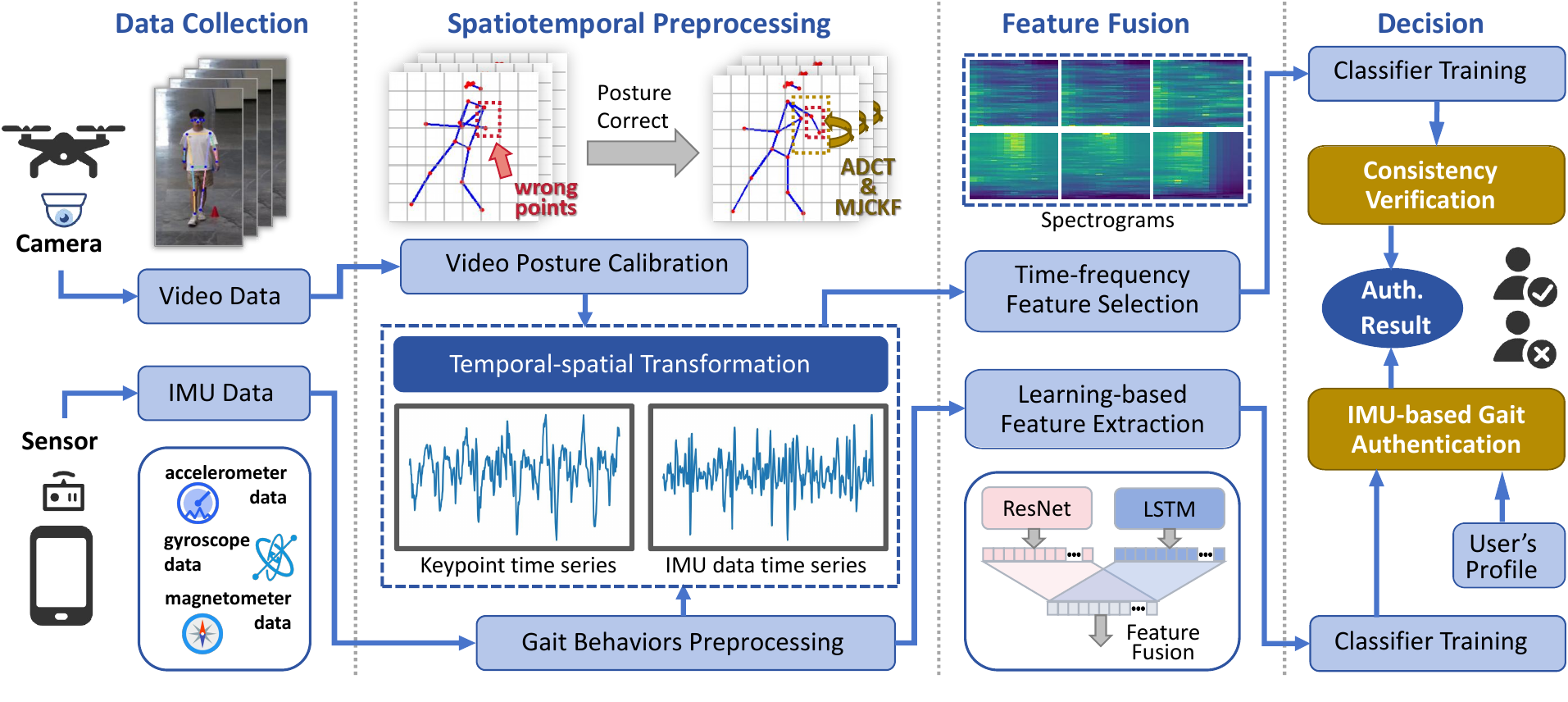} 
  \caption{\zcomment{The workflow of SyncGait.}}
  \label{fig:pattern} 
\end{figure*}

\subsection{Threat Model}
The attacker seeks to impersonate a legitimate user and deceive delivery drones into illegally delivering packages, or use malicious drones to steal users' packages. This paper focuses on several typical attack methods:


\textbf{Radio Relay Attack.} 
Radio relay attack is a sophisticated man-in-the-middle (MITM) technique in which an attacker places a relay device between the victim’s device and the legitimate terminal to intercept and forward wireless signals, facilitating fraudulent communication ~\cite{hassija2021fast}.
As illustrated in Figure \ref{fig:attacks} (a), the attacker $U'$ can employ GPS spoofing and Bluetooth beacon spoofing~\cite{jeon2018ble} to mislead the positioning information of drones $D$ and users $U$. 
At this stage, the attacker $U'$ establishes communication channels between the smartphone $S$ of the legitimate user and the legitimate drone $D$, thereby initiating the authentication process.
We assume that during authentication, the attacker can intercept and relay communications between the legitimate drone $D$ and the smartphone $S$. 
The attacker employs a malicious drone to deceive the user and obtain his IMU data, which is then relayed via the attacker's smartphone to the legitimate drone $D$, aiming to bypass SyncGait authentication.


\textbf{Device Hijacking Attack.} 
This attack assumes that the attacker can either direct access or indirect control over the user’s device, as shown in Figure \ref{fig:attacks} (b). Common methods include physical and software hijacking.
Physical hijacking – The attacker gains direct access to the user’s smartphone to execute malicious actions.
Software hijacking – The attacker exploits software vulnerabilities to elevate control over an application ~\cite{heuser2017droidauditor}. For instance, they may steal the victim’s login credentials for the delivery app ~\cite{facebook2019passwords}.
Once compromised, the attacker can clone the victim’s device environment on his own device and impersonate the victim during drone authentication. However, modifying the user’s configuration data remains restricted, as this typically requires higher privileges (\eg, the user’s personal password).

\textbf{Mimicry Attack.} 
Mimicry attack is an advanced threat targeting SyncGait, where the attacker attempts to bypass authentication by replicating the legal user’s gait patterns~\cite{muaaz2017smartphone}. As shown in Figure \ref{fig:attacks} (c), this attack requires the radio relay attack or the device hijacking attack as a prerequisite.
We assume that the attacker can obtain the victim’s gait data by covertly recording their movements in public spaces or via hidden cameras.
To refine the mimicry, the attacker may analyze video recordings and even practice in real-world scenarios, leveraging modern technologies to enhance accuracy ~\cite{fernandez2020reliability}. Through iterative adjustments, the attacker gradually approximates the victim’s gait characteristics, increasing the likelihood of bypassing authentication.

\textbf{Out-of-Scope.} 
The attacker may use robots to carry out sophisticated imitation attacks. The significant delays involved in data collection, transmission, processing, and robot response make real-time imitation challenging. Current humanoid robots, such as NAO, experience action execution delays of approximately 1.72 seconds~\cite{g2auth}, which far exceed the time it takes for humans to imitate. Additionally, their high cost, up to \$14,990~\cite{robotlab2022nao}, further raises the barriers to the feasibility of such attacks. Since these response time delays and the high cost are unlikely to be resolved soon, this study excludes the robot-based attack from consideration.

%% file: Section/4_System_Design_revised.tex
\section{System Design}

\subsection{Overview of SyncGait}
To address the challenges of sophisticated spoofing attacks and ensure stable long-range authentication in varying scenarios, SyncGait introduces a robust mutual authentication mechanism that leverages each user's unique and hard-to-replicate gait behaviors. 
By extracting and analyzing cross-modal consistency between video data captured by the drone’s onboard camera and IMU data from the user’s smartphone,
SyncGait achieves near-perfect authentication even at distances exceeding 18 m.

As shown in Figure~\ref{fig:pattern}, 
\zcomment{SyncGait consists of four primary steps: \emph{Data Collection}, \emph{Spatiotemporal Preprocessing}, \emph{Feature Fusion}, and \emph{Decision}. 
In the SyncGait process, the drone captures video data through its camera while a smartphone simultaneously records IMU sensor data reflecting the user's gait. 
The collected video data is processed through an adaptive discrete cosine transform (ADCT) combined with multi-joint cooperative Kalman filtering (MJCKF) to address critical issues of noise interference and keypoint occlusion. 
Subsequently, we introduce a proposed Temporal-spatial transformation mechanism to precisely align the IMU and posture data streams, compensating for temporal latency and spatial coordinate discrepancies. 
We perform feature computation on the aligned IMU and posture data streams to achieve robust and stable Consistency Verification. 
The preprocessed IMU gait data undergoes gait behavior characterization based on rotational angles to extract discriminative features. These features are then input into a hybrid deep learning model for deep feature extraction and fusion. Leveraging transfer learning, we achieve effective and robust IMU-based Gait Authentication. 
Both verification and authentication processes are evaluated concurrently. The drone delivery process commences only after both have been completed. }

\subsection{Data Collection}
As the user approaches the drone for mutual authentication, the smartphone held in the user's hand collects IMU data while the drone's camera records the user's gait video. The IMU data includes implicit time-series gait information from the accelerometer, gyroscope, and magnetometer, whereas the video data captures the user's visual gait information.


\hcomment{Pose detection can effectively mitigate the challenges arising from background interference and motion blur caused by long distances.
Compared to traditional small object detection to locate the smartphone in the user's hand~\cite{g2auth}}, our pose detection offers significant advantages in capturing human motion characteristics. It comprehensively captures the coordinated movements between different body parts, which is essential for accurately identifying complex gait features. 
In SyncGait, the drone captures the time-series data of the user's body keypoints in real-time and uses object detection algorithms to identify the keypoints of the hand holding the smartphone.

\subsection{Spatiotemporal Preprocessing}

This section presents a spatiotemporal preprocessing framework that addresses noise, occlusion, and misalignment in multimodal gait data. Our approach integrates a processing pipeline of adaptive filtering, multi-joint Kalman correction, and wavelet-based IMU denoising, followed by Kalman-assisted synchronization and quaternion-based alignment, to achieve robust spatiotemporal fusion.

\subsubsection{Video Posture Calibration}
\zcomment{To extract reliable human posture data from video data, we address two core challenges: noise interference and keypoint occlusion. Specifically, we propose an adaptive cutoff filtering method that dynamically optimizes gait data extraction by analyzing the spectral characteristics of the gait cycle. In addition, we mitigate inaccuracies caused by keypoint occlusion through the combination of ADCT and MJCKF.}

\textbf{Adaptive Filtering.}
\label{section:4.2.2} 
\hcomment{In this step, we first denoise the data and analyze the spectral characteristics of the time-series using the fast Fourier transform (FFT). We then apply adaptive cutoff filtering to dynamically optimize the analysis for different users.}
After obtaining the temporal-spatial information of the body keypoint data and the IMU data, we apply a wavelet denoising algorithm to enhance data accuracy and minimize noise interference. 
Additionally, we normalize the data to ensure both sets fall within the same numerical range.

By performing frequency domain analysis on the keypoint time-series, 
we can identify the principal frequency range of the user’s gait cycles.
However, since gait frequency characteristics vary with individual walking speeds and unique gait patterns, the fixed-threshold filter may not effectively adapt to the distinct gait features of different individuals.
To optimize this process, we automatically adjust the cutoff frequency of the Butterworth filter based on the frequency domain energy distribution of the user's body keypoint time-series. Specifically, we use the FFT to analyze the spectral characteristics of the time-series, determining the frequency range where the main energy components are concentrated. We then set the high-pass and low-pass frequencies of the filter to the lower and upper bounds of this range. Through this adaptive cutoff filtering method, we can dynamically optimize the processing of both the IMU data and the body keypoint data for different users.

\begin{figure}[t]
\centering
\includegraphics[width=1\linewidth]{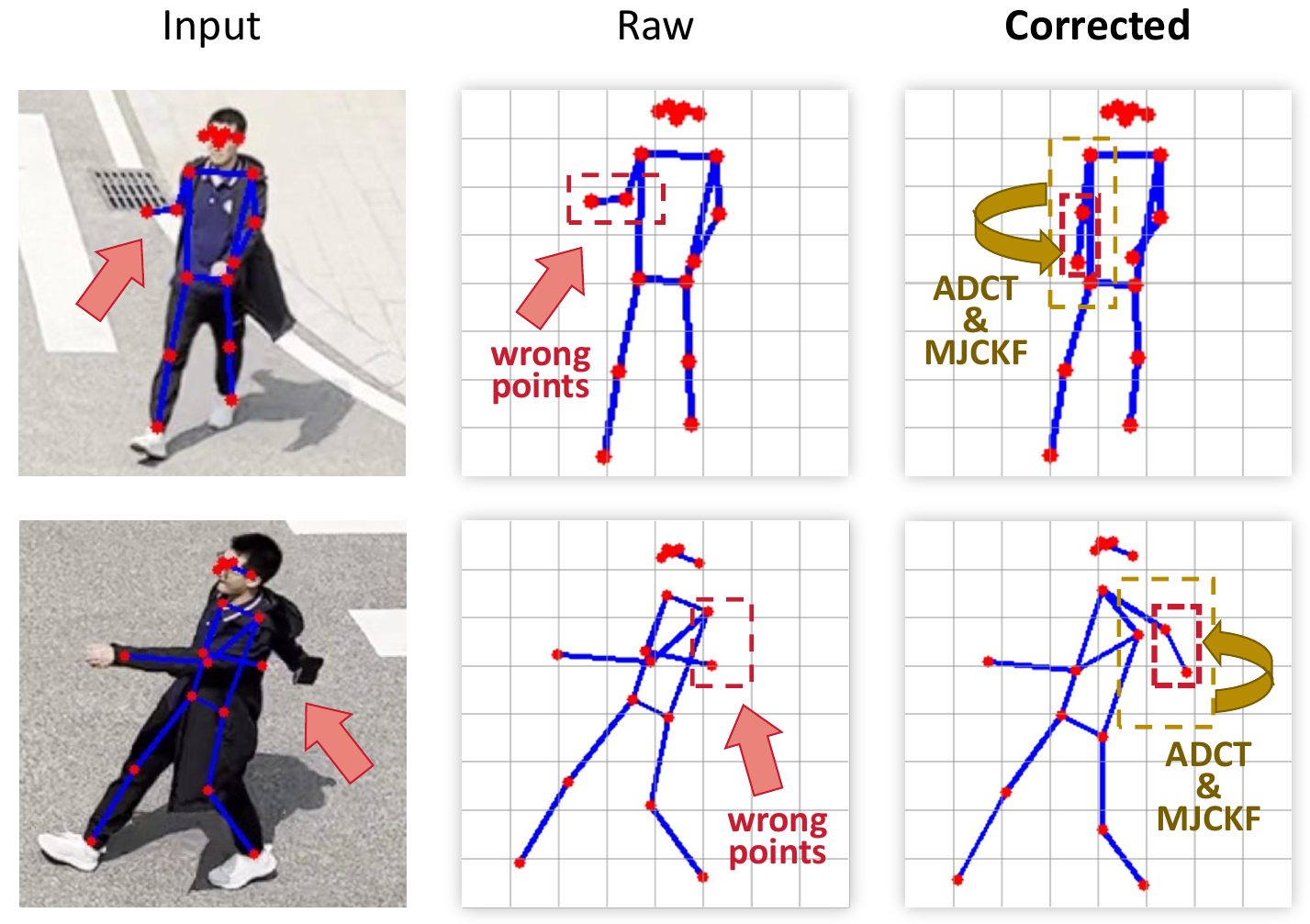}
\caption{The samples of posture calibration.}
\label{fig:PoseDCT}
\end{figure}

\textbf{Posture Correction.}
\label{section:4.2.3} 
When the drone camera is positioned at a high vertical angle or when its horizontal angle is significantly offset, the user's hands may become obscured, leading to inaccuracies in the detected keypoint positions. 
To address this issue, we exploit the fact that occlusions manifest as brief anomalies in the time-series data. By applying adaptive discrete cosine transform (ADCT) and multi-joint cooperative Kalman filtering (MJCKF), we adaptively filter out these anomalies caused by occlusion. Additionally, by incorporating information from adjacent body parts, we predict and correct the position of keypoints, effectively mitigating the impact of occlusions. As shown in Figure \ref{fig:PoseDCT}, through our processing, the positioning of the user's occluded hands can be well-corrected.

In the ADCT process, we utilize DCT to eliminate signal anomalies caused by body occlusions. Compared to traditional cutoff filters, DCT identifies intervals of signal anomalies more effectively while preserving the overall characteristics of the signal, thereby minimizing the loss of critical information~\cite{Zhao_2023_CVPR}. By retaining only the first $K$ DCT coefficients, ADCT achieves smooth signal processing. To dynamically adjust the number of retained DCT coefficients $K$, we introduce an entropy-based adaptive adjustment method. $K$ is calculated as follows:
\begin{equation}
K=\left\lfloor N  \left(f_{\text {base }}+\alpha \cdot \frac{H(X)}{\log _2 N}\right)\right\rfloor,
\end{equation}
where $f_{\text{base}}$ is the initial number of retained coefficients, $\alpha$ is an adaptive tuning parameter, $H(X)$ represents the entropy of the keypoint time-series data, and $N$ is the data length. 

Although ADCT processing effectively suppresses anomalous fluctuations in the keypoint position sequences, it can introduce signal distortion and edge effects. To mitigate this, we propose the multi-joint cooperative Kalman filtering (MJCKF) method, which integrates information from multiple adjacent keypoints into the state vector, 
effectively mitigating the signal distortion introduced by ADCT.

\subsubsection{Gait Behavior Preprocessing}

After data collection, SyncGait first performs denoising, then segments the data into gait cycles and interpolates the samples to a consistent cycle length.
When the user approaches the drone while holding a smartphone, SyncGait collects IMU data from the accelerometer, gyroscope, and magnetometer at a sampling rate of $f_s=100$Hz over a short period $t$. During each authentication session, SyncGait gathers $n$ ($n = t \times f_s$) sensor samples, with each sample comprising 9 dimensions, denoted as $(a_x, a_y, a_z, g_x,g_y, g_z,m_x, m_y, m_z)$. Here, $a$, $g$, and $m$ correspond to accelerometer, gyroscope, and magnetometer readings respectively, while $x$, $y$, and $z$ indicate the three axes of space in the phone coordinate system. 
\zcomment{To enhance data quality, we apply wavelet denoising to suppress noise in the sensor data vectors.}

\zcomment{Subsequently, we adopt an extremum-based segmentation method and apply linear interpolation to normalize each sample to a fixed length, ensuring consistency across gait cycles, which enhances accuracy and robustness compared to non-segmented data.}

\subsubsection{Temporal-spatial Transformation}

We address the temporal and spatial discrepancies between IMU gait data and drone-captured keypoint data, enabling accurate and consistent data fusion. Specifically, we propose a synchronization mechanism incorporating a Kalman filter to ensure precise temporal alignment despite network latency variations. Additionally, we introduce a gradient-aided quaternion transformation method to unify the coordinate systems of smartphone-based IMU data and camera-based human body keypoints, thereby ensuring spatial consistency.

\textbf{Temporal Transformation.} 

To ensure temporal consistency between human body keypoint data and IMU data, we employ a time-synchronization protocol~\cite{Timing-sync} and utilize a Kalman filter to address network delay fluctuations, estimating the trend in clock offset variations.
This synchronization mechanism enables precise time consistency across devices, preventing data mismatches caused by time discrepancies. Subsequently, we calculate the hand movement speed at various moments during the user’s walking process from both the IMU data and the human body keypoint data.

\begin{figure}[t]
  \centering
  \includegraphics[width=0.65\linewidth]{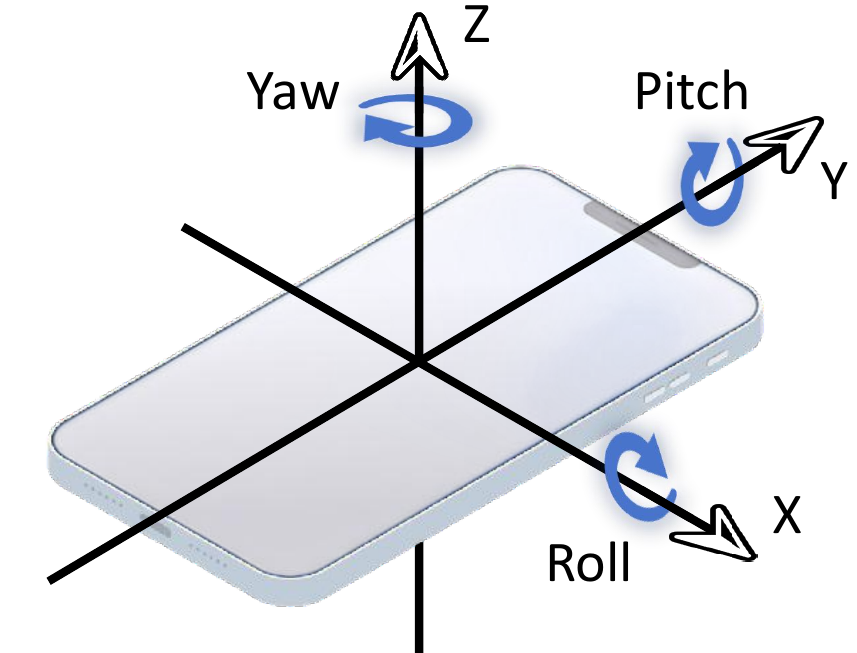} 
  \caption{The rotational movement in the coordinate system.}
  \label{fig:imu} 
\end{figure}

\textbf{Spatial Transformation.} The motion information of the user's body keypoints captured by the drone camera is referenced to a coordinate system centered on the human body, whereas the motion data measured by the IMU is referenced to a coordinate system centered on the smartphone. As the user holds the smartphone and walks to the drone, the smartphone moves with the user's hand. 
Due to the continuous change of the phone coordinate system, there is a large deviation between the motion information measured by the IMU and the user's hand motion information captured by the camera.
To eliminate these discrepancies, we first convert the IMU data from the phone coordinate system to the world coordinate system.

We propose the Gradient-Aided AHRS algorithm to improve the accuracy and robustness of quaternion-based attitude estimation.
Our method fuses the sensor data $(\boldsymbol{a_x}, \boldsymbol{a_y}, \boldsymbol{a_z}, \boldsymbol{g_x}, \boldsymbol{g_y}, \boldsymbol{g_z}, \boldsymbol{m_x}, \boldsymbol{m_y}, \boldsymbol{m_z})$, where $\boldsymbol{a}, \boldsymbol{g}, \boldsymbol{m}$ represent accelerometer, gyroscope, and magnetometer data, respectively, to obtain the orientation quaternion $\boldsymbol{q} = (\boldsymbol{q_0}, \boldsymbol{q_1},$ $ \boldsymbol{q_2}, \boldsymbol{q_3})$ By integrating gradient-based quaternion optimization and gyroscope drift compensation, we improve the accuracy and robustness of attitude estimation.

Next, the acceleration vector in the phone coordinate system $\boldsymbol{a}_{\text{phone}} = (\boldsymbol{a_x}, \boldsymbol{a_y}, \boldsymbol{a_z})$ is multiplied by the orientation quaternion $\boldsymbol{q} = ( \boldsymbol{q_{0}}, \boldsymbol{q_{1}}, \boldsymbol{q_{2}}, \boldsymbol{q_{3}} )$ and its conjugate $\boldsymbol{q}^{-1}$ to complete the conversion to the world coordinate system: $\boldsymbol{a}_{\text{world}} = \boldsymbol{q} \times \boldsymbol{a}_{\text{phone}} \times \boldsymbol{q}^{-1}$. 
Subsequently, we calculate the motion velocity from the sensor data. Setting the initial velocity at the starting time is zero, the formula for calculating the hand's motion velocity $\boldsymbol{v}_{\text{world}, t}$ at time $t$ is given by $\boldsymbol{v}_{\text{world}, t} = \boldsymbol{v}_{\text{world}, t-i} + \sum_{k=t-i}^{t} \boldsymbol{a}_{\text{world}, k} \cdot \frac{1}{f_s} \cdot i$, 
where $\boldsymbol{v}_{\text{world}, t}$ and $\boldsymbol{a}_{\text{world}, t}$ represent the velocity and acceleration of the hand at time $t$, $f_s$ is the sampling rate of the accelerometer, and $i$ is the time step.

\subsection{Feature Fusion}

In this section, we propose a feature extraction and fusion framework for user authentication using hand motion and IMU data. IMU signals and drone keypoints are projected into a shared latent space and temporally aligned to bridge modality gaps, enabling robust authentication under occlusion and noise.

\textbf{Time-Trequency Feature Selection.}
Subsequently, we extract both time-domain and frequency-domain features from the user’s hand movement data, derived from IMU and keypoint time-series signals.
After computing the normalized Fisher scores for all features, those with scores above 0.7 are selected as the optimal feature set, resulting in four time-domain features—PCC, Spearman’s rank correlation coefficient, MAE, and signal synchronization—and two frequency-domain features—coherence and spectral difference. The processed samples thus form a 6-dimensional feature vector.

\textbf{Learning-based Feature Extraction.}
\hcomment{In SyncGait, we utilize acceleration and rotational angles to characterize the user's gait behaviors, offering a more accurate representation of key rotational aspects of behaviors~\cite{wu2020liveness}.}
As illustrated in Figure~\ref{fig:imu}, the smartphone's rotational movement during gait is parameterized using Euler angles: Roll ($\phi$), Pitch ($\theta$), and Yaw ($\psi$). We compute the quaternion $\boldsymbol{q} = (q_{0}, q_{1}, q_{2}, q_{3})$ from the accelerometer, gyroscope, and magnetometer data using the method described in Section \ref{section:4.2.3}. The corresponding rotational angles are then calculated as follows:
\begin{align}
\phi &= \arctan2\left(2\left(q_{2}q_{3} + q_{0}q_{1}\right), 1 - 2\left(q_{1}^2 + q_{2}^2\right)\right) \label{eq:phi_i} \\
\theta &= \arcsin\left(2\left(q_{1}q_{3} - q_{0}q_{2}\right)\right) \label{eq:theta_i} \\
\psi &= \arctan2\left(2\left(q_{1}q_{2} + q_{0}q_{3}\right), 1 - 2\left(q_{2}^2 + q_{3}^2\right)\right) \label{eq:psi_i}
\end{align}

Ultimately, we obtain the time-series data for the hand's acceleration and rotational angles as the user approaches the drone, represented as \( (\boldsymbol{a}_x, \boldsymbol{a}_y, \boldsymbol{a}_z, \phi, \theta, \psi) \). \zcomment{We subsequently input it into a deep learning-based feature extractor constructed using transfer learning, aiming to analyze the discriminative features of gait behaviors. }

\begin{figure}[t]
  \centering
  \includegraphics[width=1\linewidth]{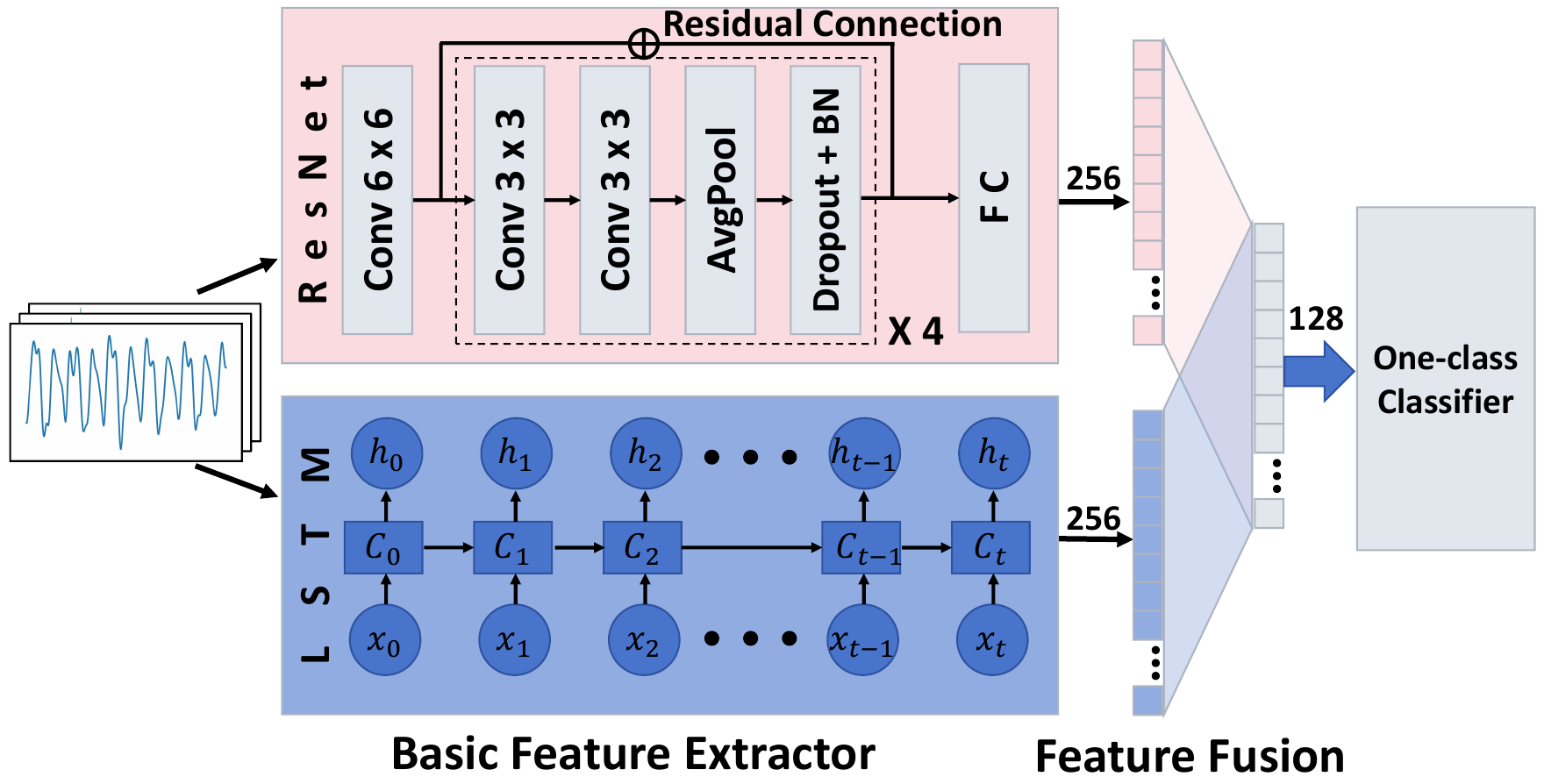} 
  \caption{A hybrid model architecture combining ResNet and LSTM networks.}
  \label{fig:network}
\end{figure}

To analyze the discriminative features of gait behaviors, we construct a deep learning-based feature extractor using transfer learning. We first pre-trained a base model on gait representation data collected from various subjects. The pre-trained model is then adapted as a general feature extractor to capture the gait characteristics of the individual users.
\zcomment{Our approach employs a hybrid architecture that integrates ResNet and LSTM networks to effectively capture both spatial and temporal characteristics of gait behaviors. The model architecture is shown in the figure~\ref{fig:network}. Specifically, a bidirectional LSTM network is introduced to model the temporal dependencies within gait sequences, while ResNet extracts spatial features from gait representations. To enhance the overall feature representation, the spatial features obtained from ResNet are concatenated with the hidden states of the LSTM along the feature dimension, facilitating a more comprehensive characterization of gait behaviors. 
Additionally, we explore alternative model architectures, including CNN-based models, and evaluate their performance, which will be discussed comprehensively in Section~\ref{section:4.2}.}

To reduce the model's dependence on the data, we employ transfer learning, using the base model as a feature extractor and connecting a one-class classifier for user authentication.
We train the base model using IMU data samples collected from 11 subjects, with each subject providing 300 samples. The batch size and number of epochs are set to 128 and 1500, respectively. After pre-training the base model, we remove the fully connected (FC) and SoftMax layers, using the model as a feature extractor.
The base model requires training only once and can then be applied to unseen subjects for feature extraction.

\subsection{Verification 
 \& Authentication}
In real-world user-drone authentication scenarios, the training dataset typically contains data only from the legitimate user. 
Therefore, \zcomment{after extracting deep features, we employ one-class classifiers for authentication~\cite{zhou2024securing}. Specifically, following the extraction of each legitimate user’s features through the time-frequency feature selection and learning-based feature extraction, we train two distinct one-class classifiers per user—one for Consistency Verification and the other for IMU-based Gait Authentication. Within SyncGait, the drone delivery process is initiated only when both verification and authentication are successful.}

We use data samples collected from 11 subjects to train the one-class classifier and data samples from another 20 subjects to evaluate multiple one-class classifiers, \eg, one-class support vector machine (OC-SVM)~\cite{scholkopf2001estimating}, centroid classifier (CC)~\cite{lovisotto2020biometric}, and decision tree (DT)~\cite{song2015decision}.
Based on performance testing Section \ref{section:5.3}, we select OC-SVM as the classifier and optimize its hyperparameters using the grid search method. 
Ultimately, the model achieves an average balanced accuracy of 99.46\% on the test set of the other 20 subjects. 
It is worth noting that the consistency validation model only needs to be pre-trained once, after which it can be transferred to unseen subjects for decision.

%% file: Section/6_Evaluation_revised.tex
\section{Evaluation}

To evaluate SyncGait, we collected abundant IMU and keypoint time-series data from participants across diverse real-world scenarios, utilizing multiple drones and smartphones. Our experiments emphasize realistic conditions, including variations in drone altitude, camera resolution, sampling rates, and authentication distances. Comprehensive metrics are employed to rigorously quantify system performance and resilience against sophisticated adversarial threats.

\begin{table*}[t]
\centering
\caption{Summary of 14 Datasets in SyncGait Evaluation}

\label{Tab:Dataset}
\renewcommand{\arraystretch}{1.1}
\renewcommand{\scriptsize}{\fontsize{8}{10}\selectfont}
\scriptsize
\setlength{\tabcolsep}{4pt}
\begin{tabular}{lcclll}

\toprule
\textbf{Dataset} & \textbf{\# Subjects} & \textbf{\# Data Points} & \textbf{Devices} & \textbf{Environment} & \textbf{Experiment / Purpose} \\
\midrule
Dataset-1  & 11  & 1,650  & Default & Default & Pre-training \\
Dataset-2  & 20  & 3,000  & Default & Default & Overall performance evaluation \\
Dataset-3A & 6   & 4,500  & Default & Distance = 6–18 m & Baseline: SyncGait \\
Dataset-3B & 6   & 6,000  & Default & Distance = 6–18 m & Baseline: facial recognition \\
Dataset-3C & —   & 1,000  & Default & 6–18 m, RSSI = –80 dBm & Baseline: distance bounding \\

Dataset-4A & 20  & 2,000 & Default & 18 m distance, open area & Radio relay attack \\
Dataset-4B & 20  & 2,000 & Default & Default & Device hijacking attack \\
Dataset-4C & 6 pairs & 1,200 & Default & Real-world mimicry setup & Mimicry attacks (relay/hijack based) \\

Dataset-5  & 6   & 3,000  & Default & 5 weeks, weekly sessions & Temporal consistency evaluation \\
Dataset-6  & 6   & 8,100  & Default & Square / Yard / Park & Environment robustness test \\
Dataset-7  & 31  & 37,200 & Default & Packet loss 0–60\%, FPS 40–60 & Transmission robustness \\
Dataset-8  & 31  & 55,800 & Default & 2.7K–720P, 10–60 Hz  & Resolution / frame rate variation \\
Dataset-9  & 15  & 12,000 & Default & Left / Right hand, holding angles & Hand-holding habit analysis \\
Dataset-10 & 10  & 12,000 & 3 phones × 4 drones & Cross-device combination (12 types) & Device generalization \\
Dataset-11 & 15  & 12,000 & Default & Altitude 3–6 m, Distance 9–30 m & Hover height / distance analysis \\
Dataset-12 & 12  & 7,200  & Default & Angle 0° / 15° / 30°, occluded / clear & Angle deviation / occlusion test \\
Dataset-13 & 6   & 1,800  & Default & Noon / Sunset / Night & Lighting condition robustness \\
Dataset-14 & 12  & 10,800 & Xiaomi Mi 9, S9+, Pixel 4 & Speeds: slow / normal / fast & Walking speed and device analysis \\
\bottomrule
\end{tabular}
\end{table*}

\subsection{Experimental Settings}

\textbf{Data collection}.
After obtaining IRB approval, we recruited 31 participants (20 males and 11 females, aged 22–45) who provided informed consent. Each participant held a smartphone and walked toward a hovering drone positioned in front of them. The smartphone continuously recorded IMU data, while the drone simultaneously captured keypoint time-series. Each walking session lasted about 15 s and was segmented into five synchronized IMU–keypoint pairs, yielding 181,250 samples in total.
Based on this data, we constructed 14 datasets to evaluate SyncGait under diverse conditions, including long-distance authentication, temporal consistency, environmental variations, device heterogeneity, lighting and speed differences, and multiple attack scenarios. A balanced 1:1 ratio of positive to negative samples was maintained by randomly selecting negative data from other participants.
Specifically, Dataset-1 is used for pre-training the feature extractor and classifier, while Dataset-2 evaluates the overall authentication performance. Dataset-3 (A–C) serves to compare SyncGait with baseline methods, including facial recognition and distance-bounding schemes. Dataset-4 (A–C) assesses the system’s robustness against real-world attack scenarios such as relay, hijacking, and mimicry attacks. Finally, Dataset-5 through Dataset-14 examine SyncGait’s performance under various physical conditions, covering temporal consistency, environmental diversity, device variation, illumination changes, and walking-speed differences, as summarized in Table~\ref{Tab:Dataset}.

\begin{figure}[t]
  \centering
  \includegraphics[width=0.75\linewidth]{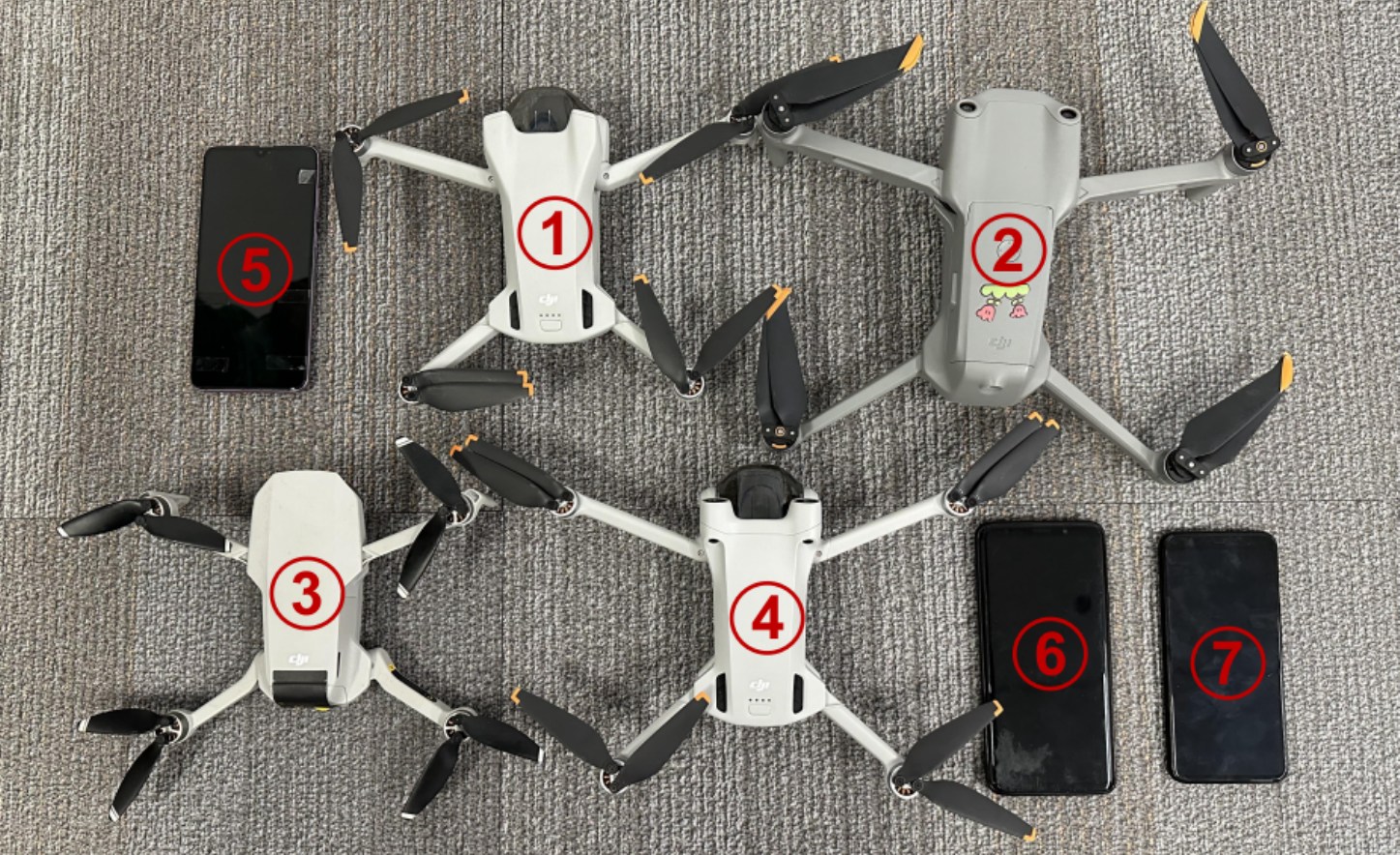} 
  \caption{Seven devices used in the experiment.} 
  \label{device} 
  \vspace{-5pt}
\end{figure}

\textbf{Default Setting}.
We used several devices for data collection, as shown in Figure~\ref{device}. The equipment used in the study includes four drones: a DJI Mini 3 drone (labeled as 1), a DJI Air 2S drone (labeled as 2), a DJI Mavic Mini drone (labeled as 3), a DJI Mini 3 Pro drone (labeled as 4), 
and three smartphones: a Xiaomi MI 9 (labeled as 5), a Samsung Galaxy S9+ (labeled as 6), and a Google Pixel 4 (labeled as 7), Gait data was captured using the camera embedded in the drone. The resolution of the drone camera was set to 2.7K, with a frame rate of 60 FPS, and the hovering altitude was maintained at 4m. We set the default authentication distance to 18m. In the evaluation section, we downsampled the data to evaluate the impact of camera resolution and frame rate on SyncGait.
To collect IMU-based gait data, we developed a prototype system on Android 9.0 (API level 28). This system accesses the device's built-in accelerometer, gyroscope, and magnetometer to collect gait data. Specifically, we set the data sampling rate to 100 Hz.
For each authentication attempt, the default recording duration is set to 4.5 seconds.

\textbf{Evaluation Metrics}.
We evaluate SyncGait using standard authentication metrics. The False Acceptance Rate (FAR) and False Rejection Rate (FRR) measure system security and usability, respectively, with lower values indicating better performance. The Equal Error Rate (EER), where FAR equals FRR, reflects overall reliability—smaller values denote higher accuracy. The Balanced Accuracy (BAC), the mean of true acceptance and rejection rates, assesses robustness under data imbalance. The Receiver Operating Characteristic (ROC) curve illustrates the trade-off between TAR and FAR, and a higher Area Under the Curve (AUC) indicates stronger discriminative capability.


\subsection{Overall Performance}
\label{section:4.2} 
In this Section, we systematically evaluate SyncGait's overall authentication performance, robustness against advanced adversarial threats, and stability over time. 

\begin{figure*}[t!]
    \begin{minipage}[t]{0.32\linewidth}
        \centering
        \includegraphics[width=\textwidth]{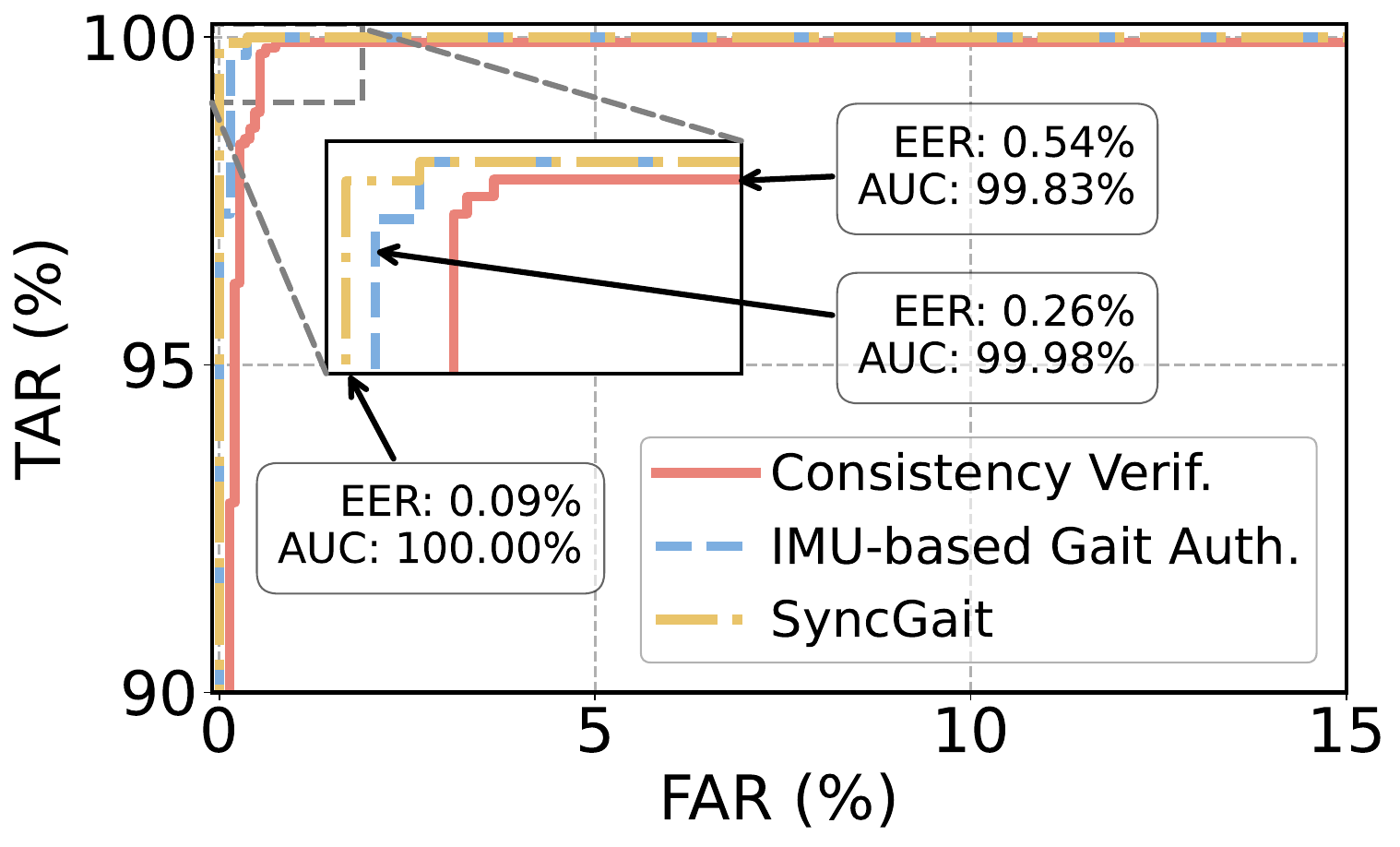} 
        \vspace{-12pt}
        \caption{ROC curves of SyncGait.\\} 
        \label{fig:6.2.1roc} 
    \end{minipage}
    \hfill   
    \begin{minipage}[t]{0.32\linewidth}
        \centering
        \includegraphics[width=\textwidth]{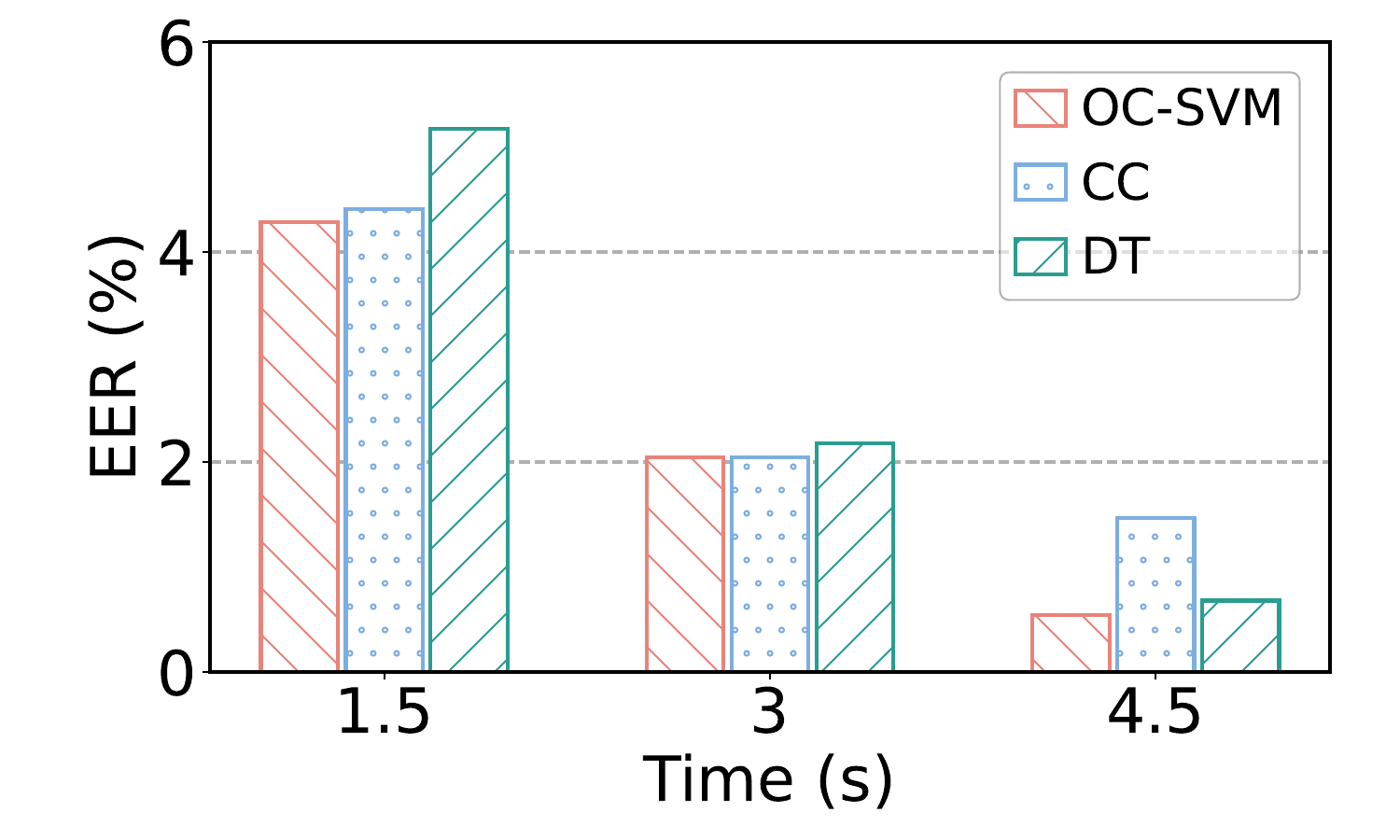}
        \vspace{-12pt}
        \caption{EERs of \hcomment{different classifier \\and} sample durations.}
        \label{fig:6.3.1}
    \end{minipage}
    \hfill
    \begin{minipage}[t]{0.32\linewidth}
        \centering
        \includegraphics[width=\textwidth]{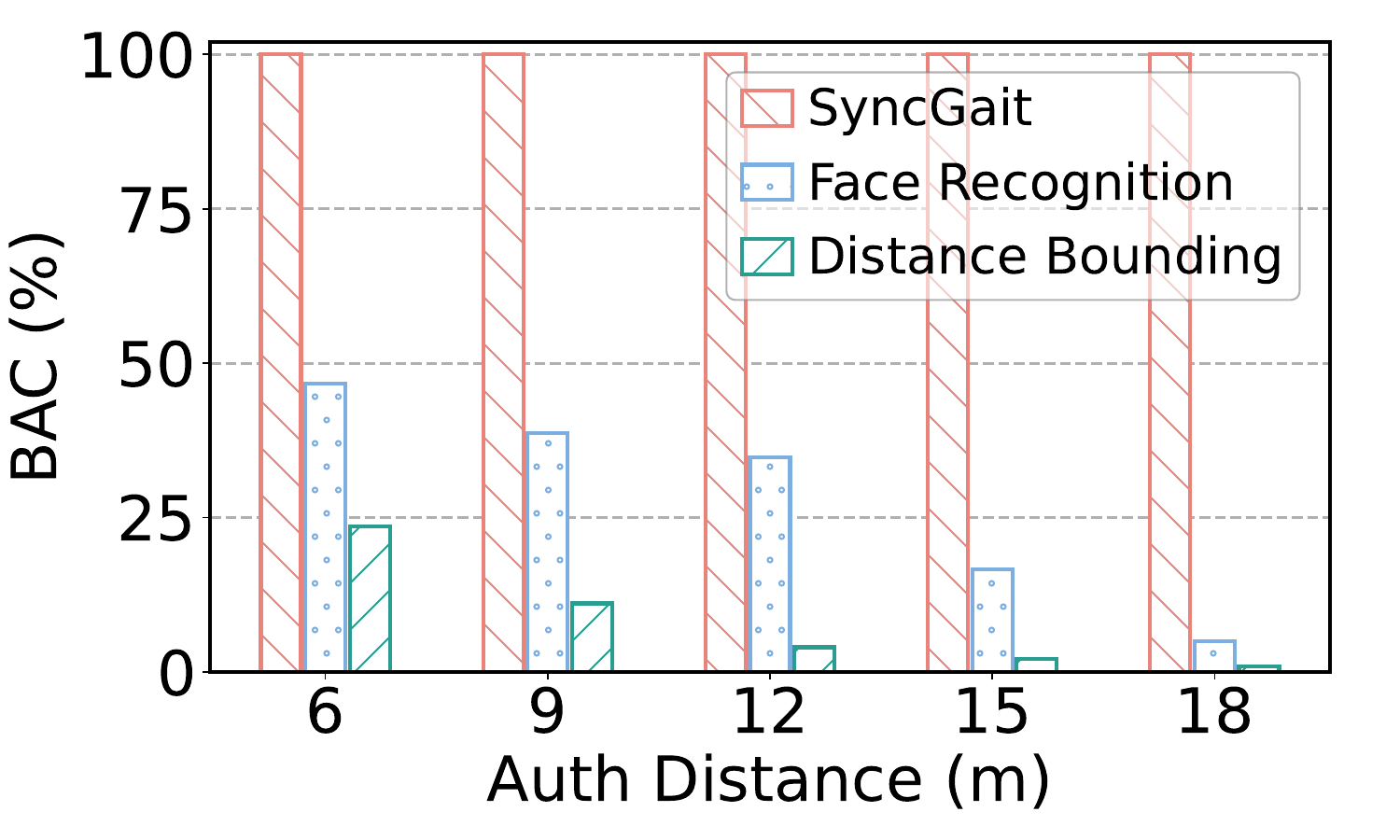}
        \vspace{-12pt}
        \caption{Baseline comparisons of\\ SyncGait.}
        \label{fig:baseline}
    \end{minipage}
\end{figure*}

\textbf{ROC Curves}.
To evaluate the overall performance of SyncGait, we trained the model using \textit{dataset-1} and assessed its effectiveness using \textit{dataset-2}. We tested the performance of \textit{Consistency Verification} module, \textit{IMU-based Gait Authentication} module, and SyncGait under one authentication attempt, respectively. 
The corresponding ROC curves are shown in Figure \ref{fig:6.2.1roc}. The \textit {Consistency Verification} module achieved an AUC of 99.83\% with an EER of 0.54\%, while the \textit{IMU-based Gait Authentication} module achieved an AUC of 99.98\% with an EER of 0.26\%. The overall performance of SyncGait demonstrated an AUC close to 100\% and an EER of 0.09\%. 
The ROC curve indicates that both modules exhibit a high TAR, nearing 100\%, while the FAR is relatively suboptimal. However, by applying a stricter criterion for positive classification through module fusion, SyncGait significantly enhances the system's ability to distinguish negative cases accurately. 
These results indicate that SyncGait exhibits superior performance in real-world scenarios.

\textbf{Impact of Classifier and Sample Duration}.
We evaluated the effects of classifier choice and sample duration on SyncGait using \textit{dataset-1} for training and \textit{dataset-2} for testing. Considering an average gait cycle of 1.5 s~\cite{zou2020deep}, we tested durations of 1.5 s, 3.0 s, and 4.5 s. As shown in Figure~\ref{fig:6.3.1}, longer samples reduced the EER of the \textit{Consistency Verification} module. Among classifiers, OC-SVM outperformed CC and DT, achieving an EER of 0.54\% at 4.5 s, which was adopted as the default setting for subsequent experiments.

\textbf{Impact of Feature Extraction Model.} 
We evaluated multiple feature extraction models for SyncGait using \textit{dataset-1} for training and \textit{dataset-2} for testing. As shown in Table~\ref{tab:model_comparison}, all models were pre-trained as feature extractors and paired with a one-class classifier. ResNet notably outperformed BiLSTM and baseCNN. Among hybrid structures, two feature concatenation methods—horizontal and vertical—were compared, with the ResNet+LSTM\_horizontal model achieving the best performance: 99.65\% accuracy, 0.59 s latency, and 45.61 MB memory usage.

\begin{table}[t!]
\caption{Performance for different feature extractors.}
\vspace{-4pt}
\label{tab:model_comparison}
\renewcommand{\scriptsize}{\fontsize{7}{10}\selectfont}
\scriptsize
\begin{center}
{
\begin{tabular}{l r r r r}
    \toprule
    Model & Accuracy & Runtime (sec) & Memory (MB) 
    \\  \midrule   
    BiLSTM & 0.9485 & 0.3473 & 2.4298 
    \\
    ResNet18 & 0.9909 & 0.5652 & 43.2203
    \\
    ResNet50 & 0.9881 & 0.6477 & 94.3470 
    \\  \hdashline    
    baseCNN & 0.9351 & 0.4671 & 1.3390 
    \\
    baseCNN+LSTM\_horiziontal$^1$ & 0.9678 & 0.5024 & 3.6156 
    \\
    baseCNN+LSTM\_vertical$^2$ & 0.9751 & 0.4967  & 3.3814 
    \\  \hdashline
    ResNet & 0.9954 & 0.5683 & 43.2812 
    \\
    \textbf{ResNet+LSTM\_horiziontal} & \textbf{0.9965} & \textbf{0.5904} 
    & \textbf{45.6124} 
    \\
    ResNet+LSTM\_vertical & 0.9930 & 0.5982 & 46.4954 
    \\

    \bottomrule
\end{tabular}
}
\end{center}
\vspace{1pt}
    \footnotesize
    \RaggedRight 
    $^1$ The outputs of both models are concatenated into a single feature vector.\
    $^2$ The first model’s output serves as the input to the second model for final feature generation.
\end{table}

\begin{figure*}[t!]
    \begin{minipage}[t]{0.32\linewidth}
        \centering   
        \includegraphics[width=\textwidth]{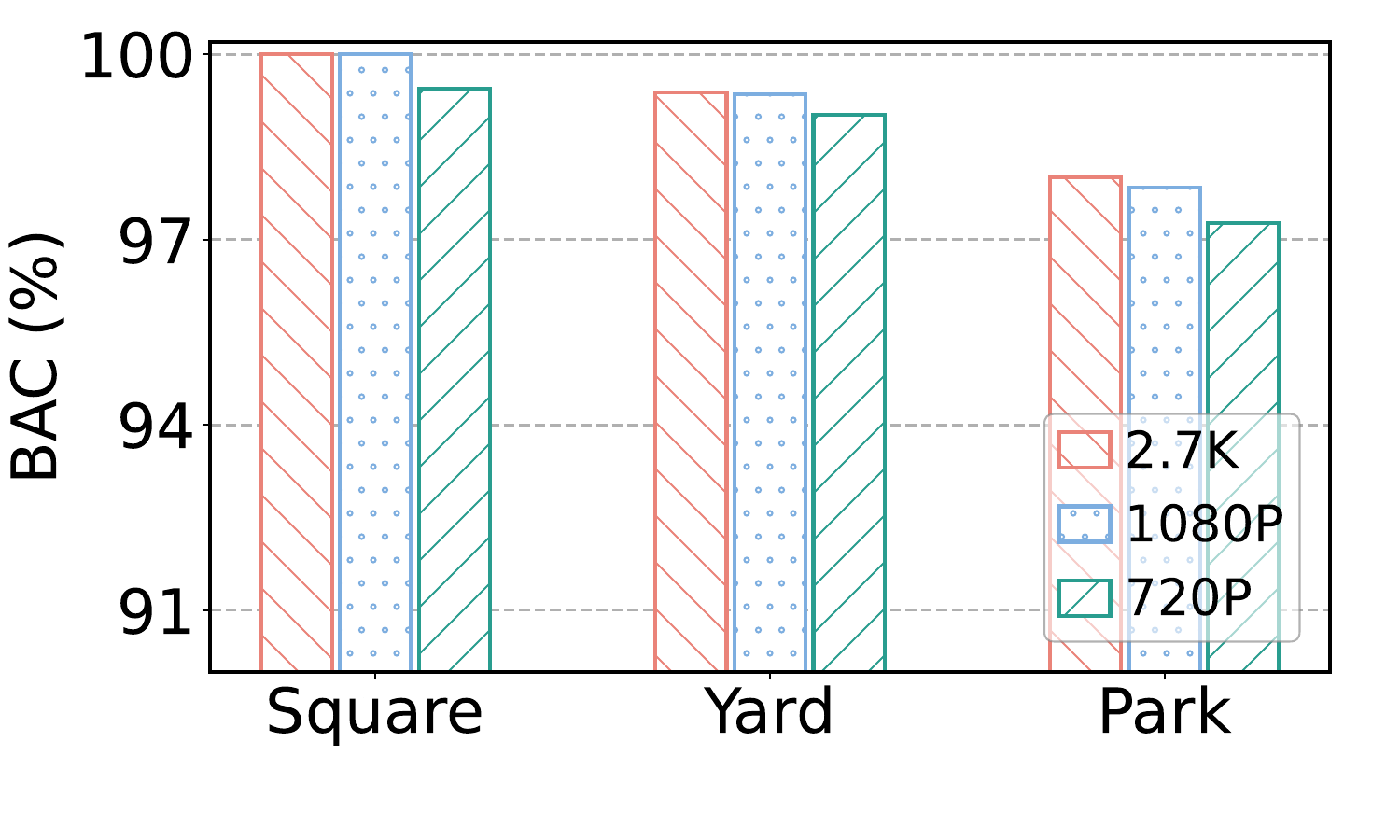}
        \caption{BACs of SyncGait under varying scenarios and video resolutions.}
        \label{fig:diff_scenario}
    \end{minipage}
    \hfill
    \begin{minipage}[t]{0.32\linewidth}
        \centering
        \includegraphics[width=\textwidth]{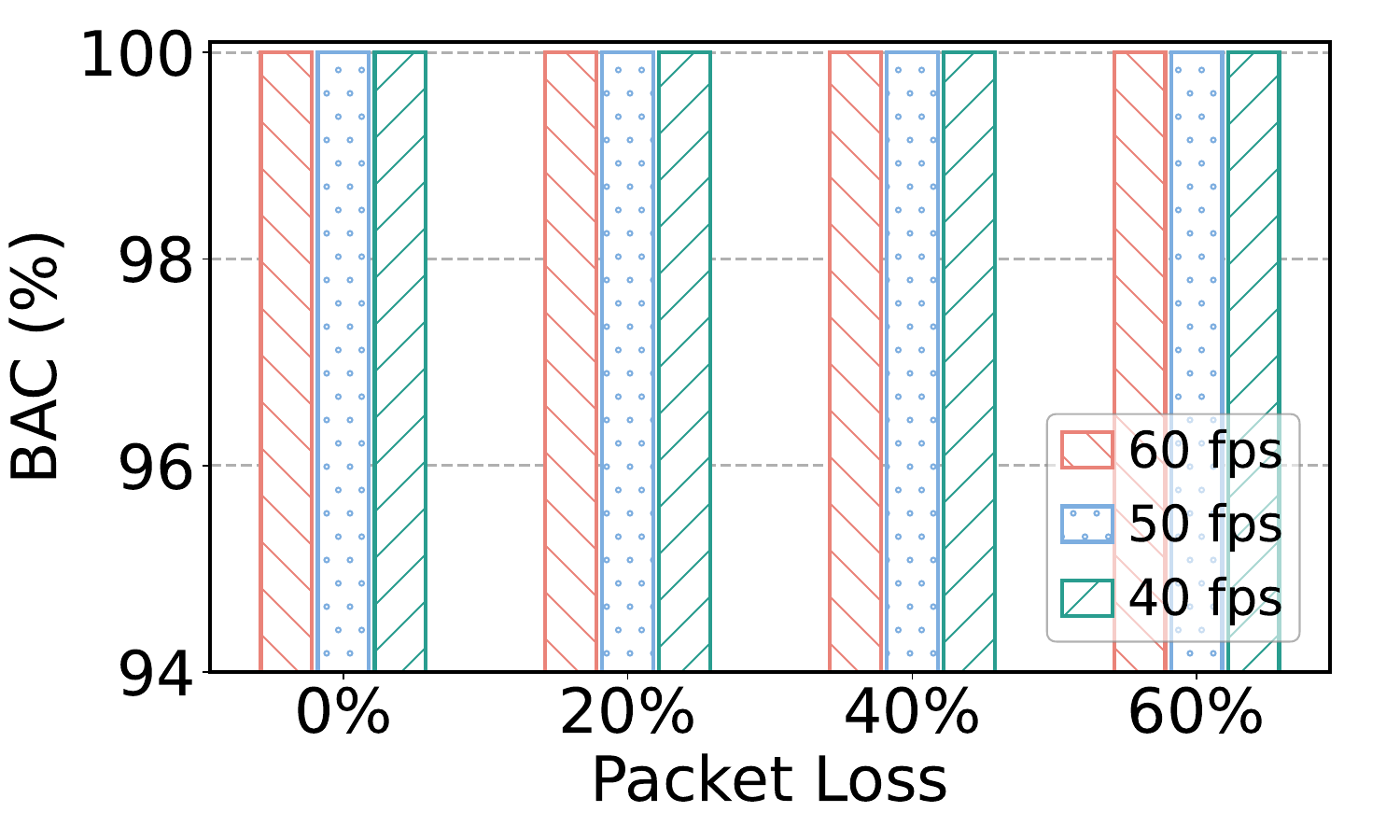}
        \caption{BACs of SyncGait under varying packet loss rates and frame rates.}
        \label{fig:packet_loss}
    \end{minipage}
    \hfill
    \begin{minipage}[t]{0.32\linewidth}
        \centering
        \includegraphics[width=\textwidth]{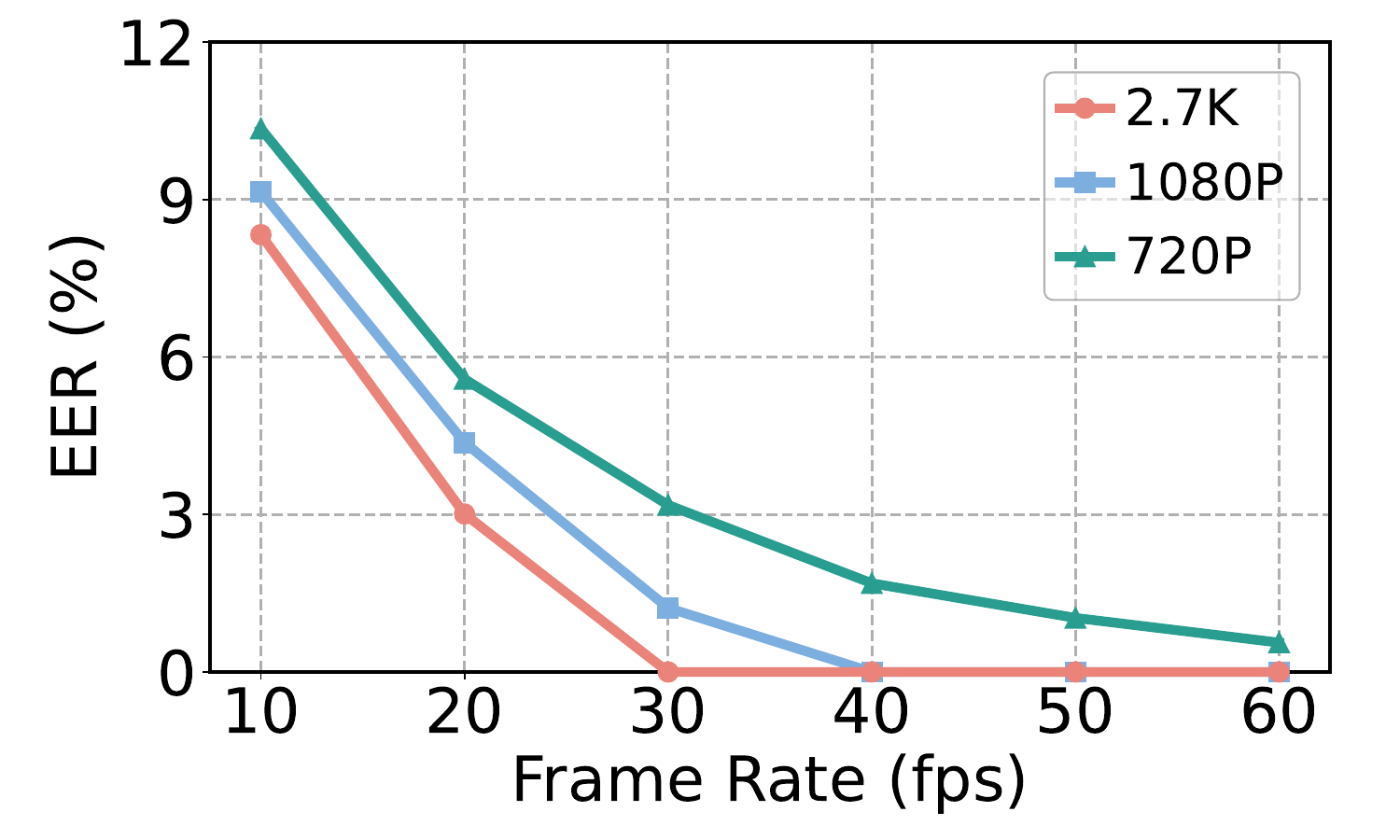}
        \caption{EERs of different video \\resolutions and frame rate.}
        \label{fig:6.3.2}
    \end{minipage}    
    \vspace{2mm}
\end{figure*}

\textbf{Baseline Comparisons}.
To evaluate SyncGait in long-distance authentication, we used \textit{dataset-3} to compare it with facial recognition~\cite{smile2auth} and distance bounding~\cite{avoine2018security} under identical settings. As shown in Figure~\ref{fig:baseline}, both baselines perform poorly at long ranges—facial recognition and distance bounding achieve 46.67\% and 23.60\% BAC at 6 m, dropping to 4.97\% and 0.90\% at 18 m, respectively. In contrast, SyncGait maintains nearly 100\% BAC across all distances, demonstrating superior robustness and reliability.
These results stem from inherent limitations of existing methods. Face recognition depends on high-resolution input, but drone cameras (\eg, 2.7K) capture insufficient facial detail at long distances, and top-down viewing angles cause distortion and occlusion. Distance bounding relies on precise round-trip time (RTT) measurements, yet consumer devices suffer from timestamp inaccuracy, Bluetooth latency, and multipath interference, making it unreliable for long-range authentication.

\textbf{FAR Against Various Attacks}.
We evaluated SyncGait’s resilience on \textit{dataset-4} against radio relay, device hijacking, and mimicry attacks, as summarized in Table~\ref{tab:attack}.
Under radio relay attacks, a malicious drone relays signals between the legitimate user and drone. The \textit{IMU-based Gait Authentication} module yields a high FAR of 99.93\%, since the IMU data originates from the genuine user, while the \textit{Consistency Verification} module achieves a low FAR of 0.78\% due to mismatched video keypoints. This demonstrates SyncGait’s strong defense capability against relay attacks. 
Under device hijacking attacks, the attacker gains control of or emulates the victim’s device to impersonate them. This bypasses the \textit{Consistency Verification} module but fails against the \textit{IMU-based Gait Authentication} due to distinctive gait patterns, yielding a FAR of only 0.09
In mimicry attacks, the attacker imitates the victim’s gait after observing their authentication process. Even with prior relay or hijacking success, the FAR remains low—0.98\% for hijacking-based and 0.80\% for relay-based mimicry—owing to the difficulty of replicating unique gait dynamics~\cite{eberz2018your} and SyncGait’s strict temporal synchronization.

\begin{table}[t!]
\renewcommand{\arraystretch}{0.8}
\renewcommand{\scriptsize}{\fontsize{8}{10}\selectfont}
\scriptsize
\setlength{\tabcolsep}{5pt}
    \centering
    \caption{FAR of various attacks.}
    \vspace{-4pt}
    \begin{tabular}{c c c c c}
    \toprule
    Attack type 
    & \makecell{\textit{Consistency} \\\textit{Verification}} 
    & \makecell{\textit{IMU-based Gait} \\\textit{Authentication}} 
    & \textbf{SyncGait} \\
    \midrule
    \makecell{Radio relay \\attack} & 0.78\% &99.93\% & \textbf{0.78\%} \\ 
    \makecell{Device hijacking  \\attack} & 99.73\% & 0.09\% & \textbf{0.09\%} \\
    \makecell{Relay-based \\mimicry attack} & 0.80\% & 99.94\% & \textbf{0.80\%} \\
    \makecell{Hijacking-based \\mimicry attack} & 99.75\% & 0.82\% & \textbf{0.82\%}  \\
    \bottomrule     
    \end{tabular}
    \label{tab:attack}
\end{table}

\begin{figure}[t]
        \centering
        \includegraphics[width=0.7\linewidth]{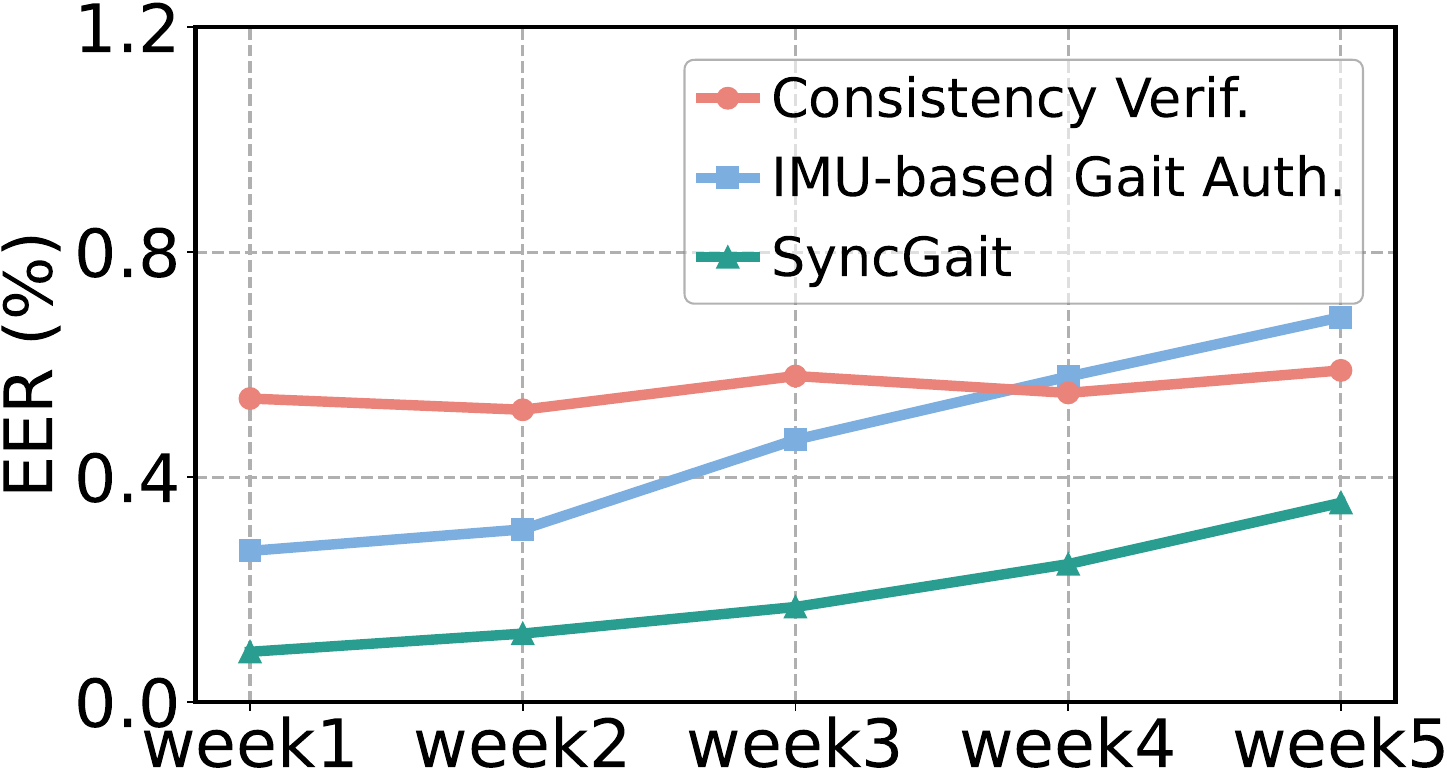}
        \caption{EERs of SyncGait over different time periods.}
        \label{fig:6.2.3}
     \end{figure}

\textbf{Performance Over Time}.
To assess the robustness of SyncGait over time, we trained the authenticators on \textit{dataset-1} and evaluated them on \textit{dataset-5}. Figure~\ref{fig:6.2.3} presents the corresponding EERs of SyncGait across the five weeks.
We can observe that the EERs of both the \textit{Consistency Verification} module and the \textit{IMU-based Gait Authentication} module fluctuated very little across several weeks. Especially, the \textit{Consistency Verification} module demonstrates particularly stable performance over time.
The slight variations in the EER of SyncGait were primarily influenced by the \textit{IMU-based Gait Authentication} module. To further enhance consistency over time, we can fine-tune the gait authentication model using recently authenticated positive samples.

\begin{figure*}[t!]
    \begin{minipage}[t]{0.32\linewidth}
        \centering
        \includegraphics[width=\textwidth]{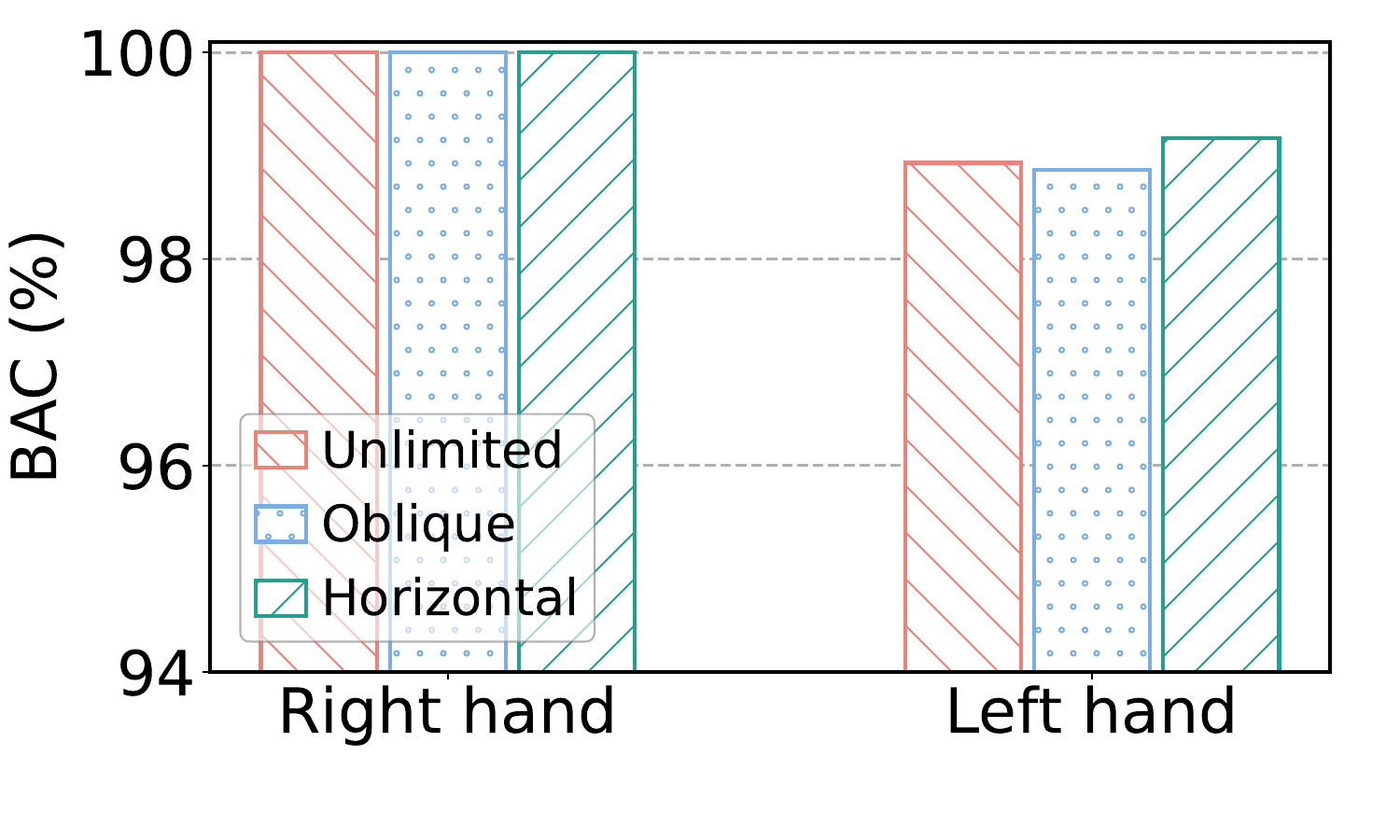}
        \caption{BACs of different phone \\holding habits.}
        \label{fig:6.3.3}
    \end{minipage}
    \hfill
    \begin{minipage}[t]{0.32\linewidth}
        \centering
        \includegraphics[width=\textwidth]{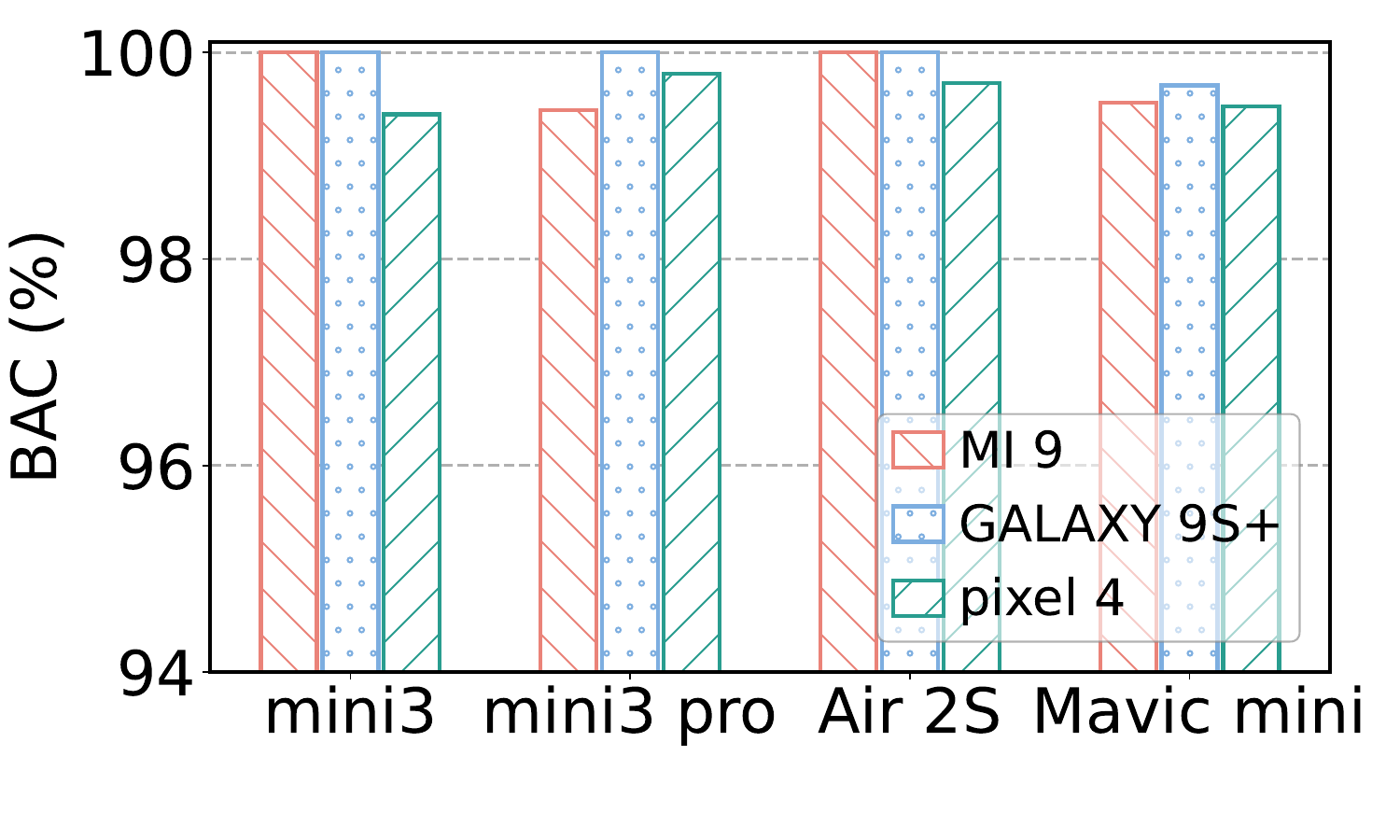}
        \caption{BACs of different phones \\and drones.}
        \label{fig:6.3.4}
    \end{minipage}
    \hfill
    \begin{minipage}[t]{0.32\linewidth}
        \centering
        \includegraphics[width=\textwidth]{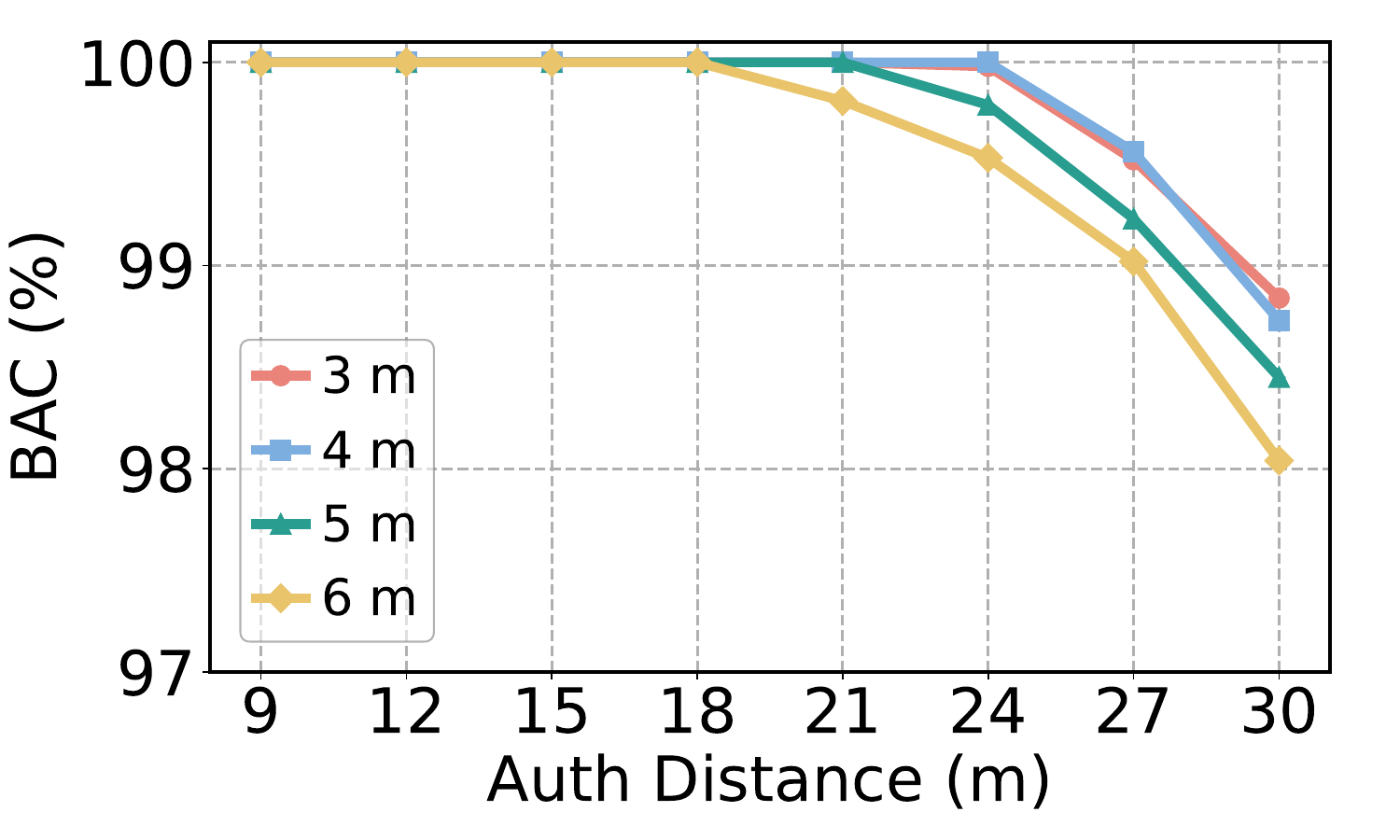}
        \caption{BACs of different hovering \\heights and distances.}
        \label{fig:6.3.5}
    \end{minipage}    
    \vspace{2mm}
\end{figure*}



\subsection{Reliability Analysis}
\label{section:5.3} 
In this section, we evaluate SyncGait under diverse real-world conditions, including environmental noise, data loss, device diversity, user habits, and visual degradation, \etc. Results show that SyncGait consistently maintains reliable authentication performance, demonstrating strong robustness and adaptability in general operational environments.

\textbf{Impact of Authentication Environment}.
We evaluated SyncGait’s robustness in complex environments using \textit{dataset-6}, which includes three representative scenarios—square, yard, and park—under three camera resolutions: 2.7K, 1080P, and 720P. The square provides an open, obstacle-free setting; the yard introduces irregular user trajectories and obstacles; and the park features frequent occlusions and dense pedestrian activity. As shown in Figure~\ref{fig:diff_scenario}, SyncGait achieves BACs of 100\%, 99.39\%, and 98.01\% in the square, yard, and park, respectively, demonstrating strong adaptability to environmental complexity. Even at lower video resolutions, performance remains stable due to the robustness of the proposed Posture Calibration module.

\textbf{Impact of Packet Loss}.
Bluetooth-based transmission is prone to packet loss, especially over long distances, challenging reliable remote authentication. Using \textit{dataset-7}, we evaluated SyncGait under packet loss rates of 0\%–60\% and frame rates of 60–40 fps. As shown in Figure~\ref{fig:packet_loss}, SyncGait maintains nearly 100\% BAC across all conditions, demonstrating strong robustness. This stability stems from the Temporal Transformation algorithm, which filters misaligned IMU–video pairs to ensure temporal consistency under adverse transmission conditions.
\begin{figure*}[t!] 
    \begin{minipage}[t]{0.32\linewidth}
        \centering
        \includegraphics[width=\textwidth]{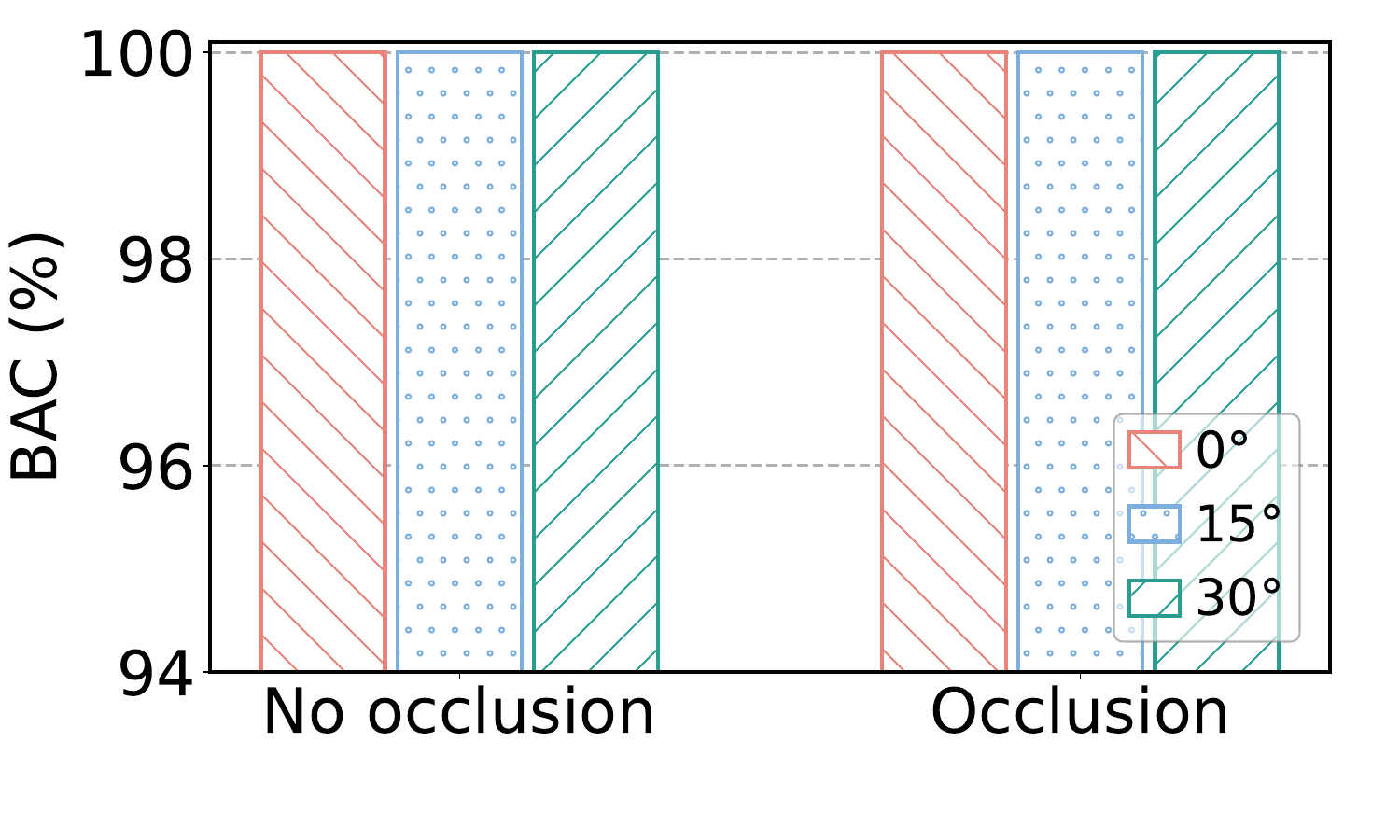}
        \caption{BACs of different angle \\deviations and occlusion conditions.}
        \label{fig:6.3.6}
    \end{minipage}
    \hfill
    \begin{minipage}[t]{0.32\linewidth}
        \centering
        \includegraphics[width=\textwidth]{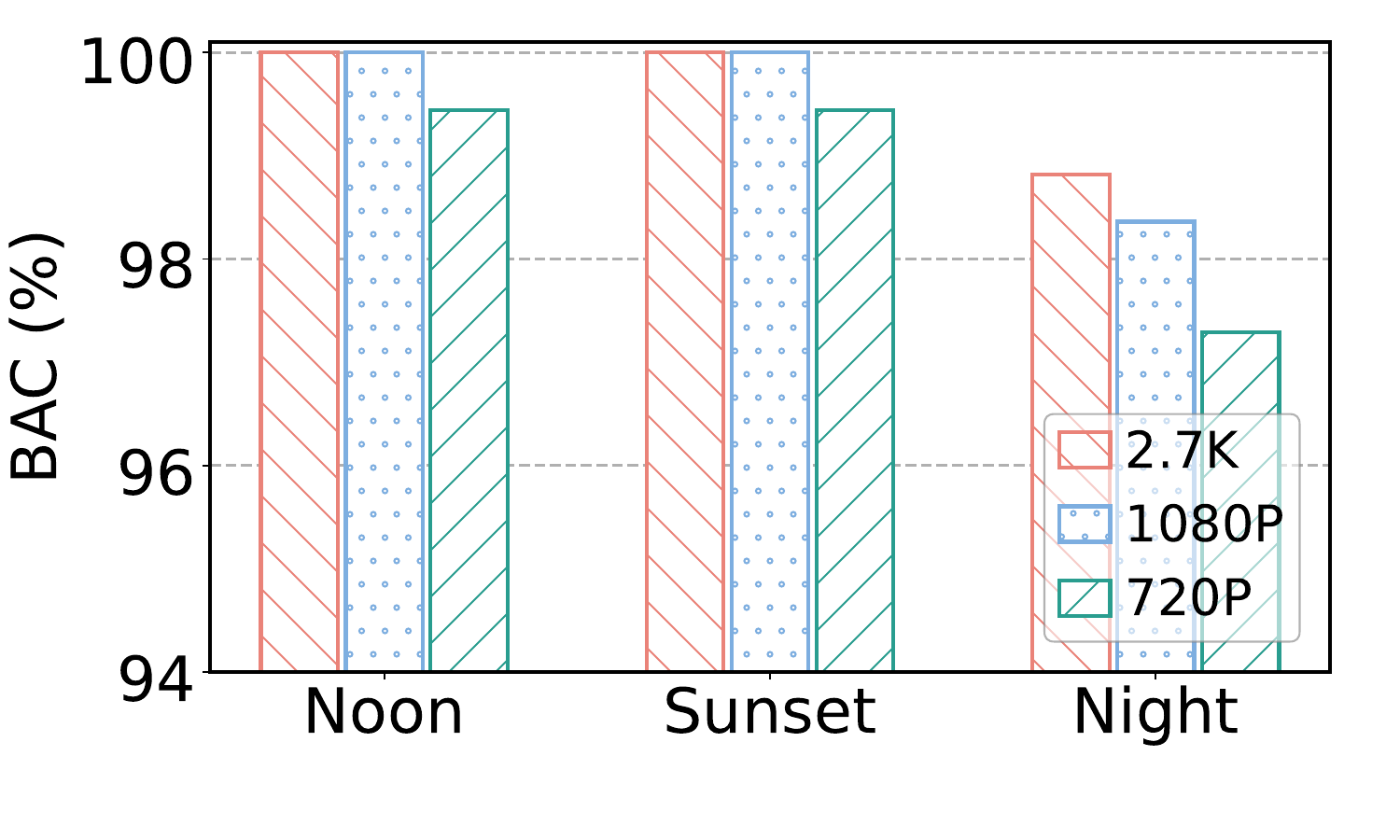}
        \caption{BACs of lighting conditions \\and video resolutions.}
        \label{fig:6.3.7}
    \end{minipage}
    \hfill
    \begin{minipage}[t]{0.32\linewidth}
        \centering
        \includegraphics[width=\textwidth]{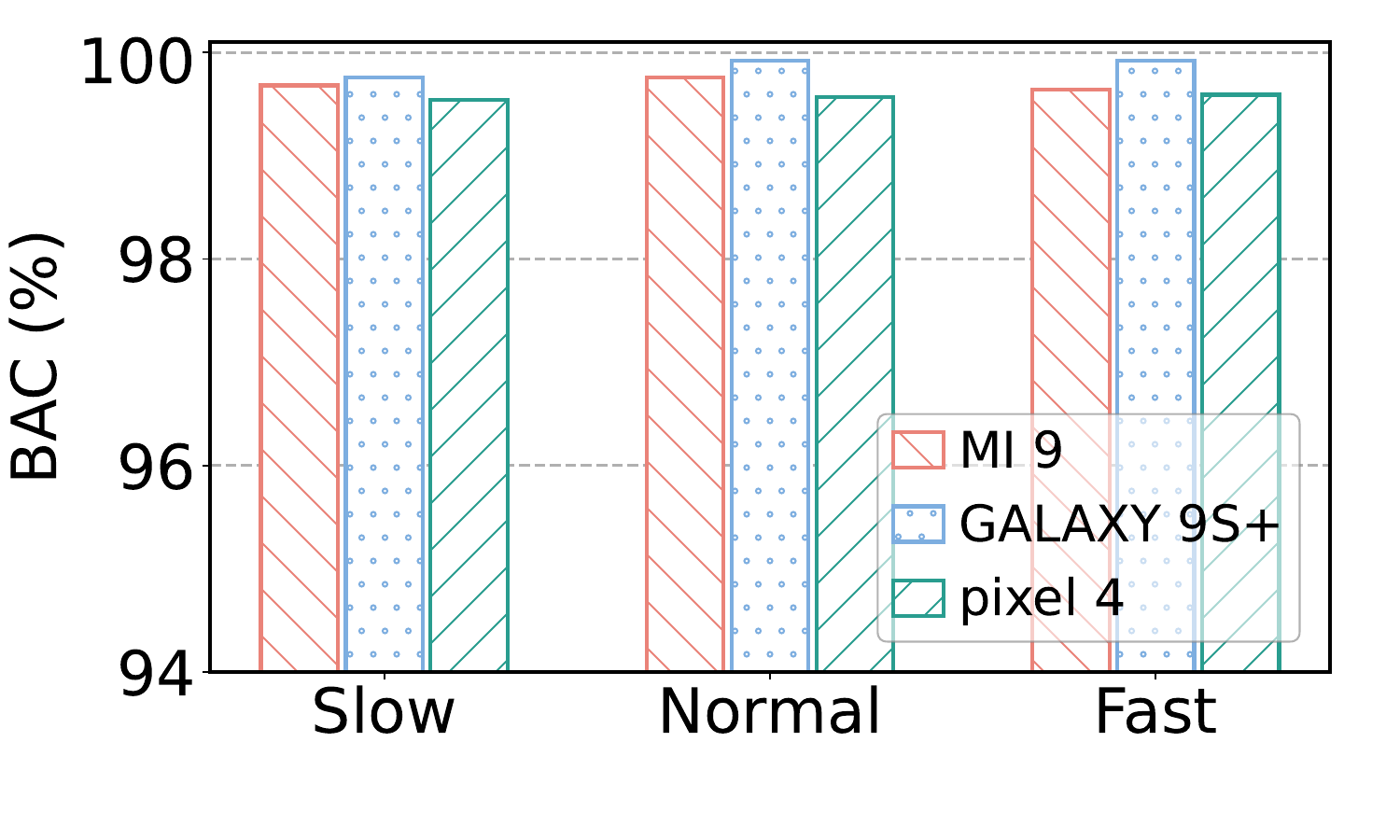}
        \caption{BACs of walking speeds and phones.}
        \label{fig:6.4.3}
    \end{minipage}
\end{figure*}

\textbf{Impact of Video Setting}.
We evaluated the impact of video frame rate and resolution on SyncGait using \textit{dataset-8}. As shown in Figure~\ref{fig:6.3.2}, the EER decreases sharply as the frame rate increases, indicating that higher temporal resolution significantly improves authentication accuracy. In contrast, the influence of video resolution (2.7K, 1080P, 720P) is relatively minor, as all three exhibit similar trends. Even at lower resolutions, SyncGait maintains stable performance, while lower frame rates—particularly below 30 fps—lead to a noticeable rise in EER due to the reduced availability of time–frequency gait features.

\textbf{Impact of Phone Holding Habit}.
This experiment evaluates the performance of SyncGait under different hand-holding habits by using \textit{dataset-9}. Figure \ref{fig:6.3.3} demonstrates the BACs under different phone-holding postures with the right/left hand. The results show that the BAC is significantly higher when the phone is held in the right hand compared to the left hand. This is because most users are accustomed to holding their phones with the right hands. Uncommon holding methods can lead to a slight decrease in authentication performance. \hcomment{The results indicate that the phone-holding posture has minimal impact on the system performance.}

\textbf{Impact of Device}.
To evaluate the impact of different devices on SyncGait, we used \textit{dataset-10}. Figure \ref{fig:6.3.4} illustrates the performance of SyncGait across various drone and smartphone combinations. 
The results suggest that, except for the Mavic Mini, the performance of SyncGait across drone devices exhibits minimal variation. This disparity is likely due to the lower video resolution of the Mavic Mini. 
The slightly lower BAC for the Pixel 4 is due to the small error in IMU data, likely caused by the device's age. Overall, the experimental results indicate that the impact of different devices on SyncGait is negligible.

\textbf{Impact of Hovering Height and Distance}.
In this experiment, we investigate the influence of drone hovering height and authentication distance on SyncGait by using \textit{dataset-11}. As shown in Figure \ref{fig:6.3.5}, when the drone's hovering height is 3m or 4m, The effect of SyncGait is almost the same. 
As the hovering height increases further, the BAC shows a slight decline. Additionally, when the authentication distance exceeds 24m, the BAC of SyncGait begins to gradually decrease due to reduced accuracy in locating the user’s body keypoints.
Therefore, we recommend setting the drone's hovering height to 4m and initiating mutual authentication at a distance of 24m.

\textbf{Impact of Angle Deviation and Body Occlusion}.
We evaluated SyncGait under angle deviation and body occlusion using \textit{dataset-12}. As shown in Figure~\ref{fig:6.3.6}, the BAC remains near 100\% at deviation angles of 0°, 15°, and 30°. Both environmental and clothing-induced occlusions have minimal impact, owing to the proposed posture correction algorithm, which adaptively filters anomalies caused by angle variations and occlusions.

\textbf{Impact of Lighting Condition}.
We investigate the influence of lighting conditions on the BAC of SyncGait via using \textit{dataset-13}. As demonstrated in Figure \ref{fig:6.3.7}, the reduced sunlight at noon and sunset does not affect the performance of SyncGait. Under night conditions, where only dim lighting is available at the experimental site, the module's performance shows a slight decline due to the further degradation of image quality in low-light environments. 
Nevertheless, even in poorly lit night conditions, SyncGait maintains a high level of BAC, demonstrating its strong adaptability to low-light environments.

\textbf{Impact of Walking Speed and Smartphones}. 
We used \textit{dataset-14} to evaluate the impact of different  walking speeds and smartphones on the performance of SyncGait. As shown in Figure \ref{fig:6.4.3}, the performance of Pixel 4 is
slightly worse than those of MI 9 SE and Galaxy S9+. This may be because its IMU sensor version is outdated and has been used for a long time. Besides, walking speed has little impact on the BAC.
These results indicate that the \textit{IMU-based Gait Authentication} module demonstrates high robustness and consistency across various devices and walking speeds.